\documentclass[%
 reprint,
 amsmath,amssymb,nolongbibliography,
 aps,
]{revtex4-2}

\usepackage{graphicx}% Include figure files
\usepackage{dcolumn}% Align table columns on decimal point
\usepackage{bm}% bold math
\usepackage{amsmath}
\usepackage{mathtools}
\usepackage{color}
\usepackage{subcaption}

\usepackage[capitalize]{cleveref}   %for \cref
\newcommand{\eq}[1]{\begin{align}#1\end{align}}

\begin{document}

\preprint{APS/123-QED}

\title{Thermodynamics of deterministic finite automata operating locally and periodically}% Force line breaks with \\

\author{Thomas E. Ouldridge}
 \email{t.ouldridge@imperial.ac.uk}
\affiliation{%
Imperial College Centre for Synthetic Biology and Department of Bioengineering, Imperial College London, London, SW7 2AZ, UK
}%

\author{David H. Wolpert}%
 \email{david.h.wolpert@gmail.com}
\affiliation{%
 Santa Fe Institute, Santa Fe, NM, 87501, USA \\
 Complexity Science Hub, Vienna, Austria \\
 Arizona State University, Tempe, AZ, USA \\
 International Center for Theoretical Physics, Trieste, Italy
}%

\date{\today}% It is always \today, today,

\begin{abstract}
Real-world computers have operational constraints that cause nonzero entropy production (EP). In particular, almost all real-world computers are ``periodic'', iteratively undergoing the same physical process; and ``local", in that subsystems evolve whilst physically decoupled from the rest of the computer. These constraints are so universal because decomposing a complex computation into small, iterative calculations is what makes computers so powerful. We first derive the nonzero EP  caused by the locality and periodicity constraints for deterministic finite automata (DFA), a foundational system of  computer science theory. We then relate this minimal EP to the computational characteristics of the DFA. We thus divide the languages recognised by DFA into two classes: those that can be recognised with zero EP,  and those that necessarily have non-zero EP. We also demonstrate the thermodynamic advantages of implementing a DFA with a physical process that is agnostic about the inputs that it processes.
\end{abstract}

\maketitle

\section{Introduction}

Szilard, Landauer and Bennett emphasized that computations have thermodynamic properties \cite{Szilard1964,Landauer1961,Bennett1982}. Lately, this insight has been enriched by stochastic thermodynamics \cite{Seifert2012,Parrondo2015,Wolpert2019stoch}, allowing rigorous analysis of computation far from equilibrium. Recent results include: a trade-off between 
the minimal amounts of ``hidden" memory and the minimum number
of discrete time-steps required to implement a 
given computation using a continuous-time Markov chain (CTMC) \cite{Wolpert2019}; the excess thermodynamic costs when the distribution of inputs to a computation does not match an optimal distribution \cite{Wolpert2020,Kolchinsky2021,Kolchinsky2017,Riechers2021} or involves statistical coupling between physically unconnected computational variables \cite{Boyd2018,Wolpert2020,Wolpert2019stoch,Wolpert2020uncertainty}; and  results on the thermodynamics of systems implementing loop-free circuits \cite{Wolpert2020}, Turing machines \cite{Brittain2021,Kolchinsky2020,strasberg2015thermodynamics}, and Mealy  machines \cite{Wolpert2019,mandal2012work,boyd2016identifying,Brittain2019}.

These analyses use minimal physical descriptions of the computations performed by the abstract constructs of computer science theory~\cite{arora2009computational,sipser1996introduction}.
Some recent work has instead probed the thermodynamics of certain types of  hardware, such as CMOS-based electronic circuits \cite{Gao2021,Freitas2021}. However, there exist practical constraints on physical computation that are not specified by the overall computation performed, but which are nonetheless relevant beyond a particular type of hardware. The thermodynamic costs of these constraints are not resolved by either of the approaches above, although their consequences can be significant \cite{Kolchinsky2021work}.

Accordingly, we ask: which kinds of thermodynamic costs necessarily arise when implementing
a computation using a physical system \textit{solely due to 
constraints that seem to be shared by all real-world physical systems that implement
digital computation}?
To begin to investigate this issue, here
we consider the minimal entropy production (EP) that arises
due to two ubiquitous constraints on real-world 
digital computers. First, the vast majority of 
modern physical  computers are periodic: they implement the same physical process at each iteration (or clock cycle) of the computation.
%that perform digital computation are synchronously clocked, i.e., they
%are ``periodic'',
% can be viewed as a computationally weaker version of Turing machines,
% % DFA are cousins of Turing machines: 
% in that both 
%applying the exact same physical physical process to implement
%each iteration of the computational system.
Second, all modern physical systems that perform digital computation
are ``local", {\it i.e.}, not all physical variables that are statistically coupled are also physically coupled when the system's state updates. Ultimately, the
reason that this constraint is imposed in
% universality reflects the fact that the success of
both abstract models of computation and real world computers is
that it allows us to break
% built upon breaking 
down complex computations into simple, iterative logical steps. %Third, the physical units that change their state during any update of the computer's state are not designed specifically to minimize EP \textit{for the precise actual distribution} of inputs they will receive during that update. 

In this work we explore how and when operating under these constraints imposes lower bounds on the EP of a computation modeled as a CTMC, regardless of any other details about how the computation is performed (equivalent results apply even in a quantum setting~\cite{Kolchinsky2021}). Taken
together, the constraints impose necessary EP through mismatch cost \cite{Wolpert2020,Kolchinsky2021,Kolchinsky2017,Riechers2021}  of two types: ``modularity'' cost  \cite{Boyd2018,Wolpert2020,Wolpert2019stoch,Wolpert2020uncertainty}, and what we call ``marginal'' mismatch cost. Both types of mismatch
cost have been identified in the literature as possibly causing EP in any given physical process; here we argue that they are in fact inescapable in complex computations. In particular, we demonstrate their effects for
one of the simplest nontrivial types of computer,
{\it deterministic finite automata} (DFA). 

% Accordingly, we ask: which thermodynamic costs necessarily arise when implementing
% a computation \textit{solely due to 
% constraints that are shared by all real-world physical systems that implement
% digital computation}?
% Specifically, we consider the minimal entropy production (EP) that arises
% due to two ubiquitous constraints on real-world 
% digital computers: periodicity, or implementing the same physical process at each iteration (or clock cycle); and  locality, or the fact that not all physical variables that are statistically coupled are also directly coupled physically when an update is performed. The universality of these constraints reflect the fact that the success of both abstract models of computation and real world computers is built upon breaking down complex computations into simple, iterative logical steps.  

% These constraints impose necessary EP through ``mismatch'' \cite{Wolpert2020,Kolchinsky2021,Kolchinsky2017,Riechers2021}  and ``modularity'' \cite{Boyd2018,Wolpert2020,Wolpert2019stoch,Wolpert2020uncertainty} costs, regardless of anyt further details; equivalent results apply even in a quantum setting~\cite{Kolchinsky2021}. Although both mismatch and modularity costs have been identified in the literature, here we argue that they are particularly relevant in practical implementations of complex computations, due to locality and periodicity constraints. Moreover, we analyse the resultant minimal EP for
% one of the simplest nontrivial types of computer,
% {\it deterministic finite automata} (DFA).

DFA have important applications in
 the design of modern compilers,
 as well as text searching and editing tools~\cite{thain2019introduction}. 
They are also foundational in computer science theory, at the foot of the Chomsky hierarchy \cite{Hopcroft1979,Moore2019}, below push-down 
automata~\cite{sipser1996introduction} and Turing machines~\cite{arora2009computational,livi08,grunwald2004shannon}.
These properties makes DFA particularly well-suited for an initial
study of the consequences of locality and periodicity in computational
systems. We thus take the first step towards investigating the thermodynamic consequences of locality and periodicity in all the computational machines of computer science theory.

We next introduce our modelling approach and key definitions.  We subsequently outline the general consequences of locality and periodicity for arbitrary computations, in the form of a strengthened second law. Having discussed these strengthened second laws, we then derive specific expressions for constraint-driven EP in DFA, and
explore how DFA could be designed to minimize the expected and worst-case costs that result. 
Next,
we analyse how this EP relates to the underlying computation performed; surprisingly, the most compact DFA for a given language is generally neither especially thermodynamically efficient nor inefficient. Finally, we 
consider
regular languages, i.e., the sets of strings such that
every string in the set can be recognized by some DFA. We
show that such languages can be divided into a class that is thermodynamically costly for a DFA to recognise, and a class that is inherently low-cost.

\section{Results}
\subsection{Deterministic Finite Automata}
\label{sec:computational}

\begin{figure}
    \centering
    \includegraphics[width=8cm]{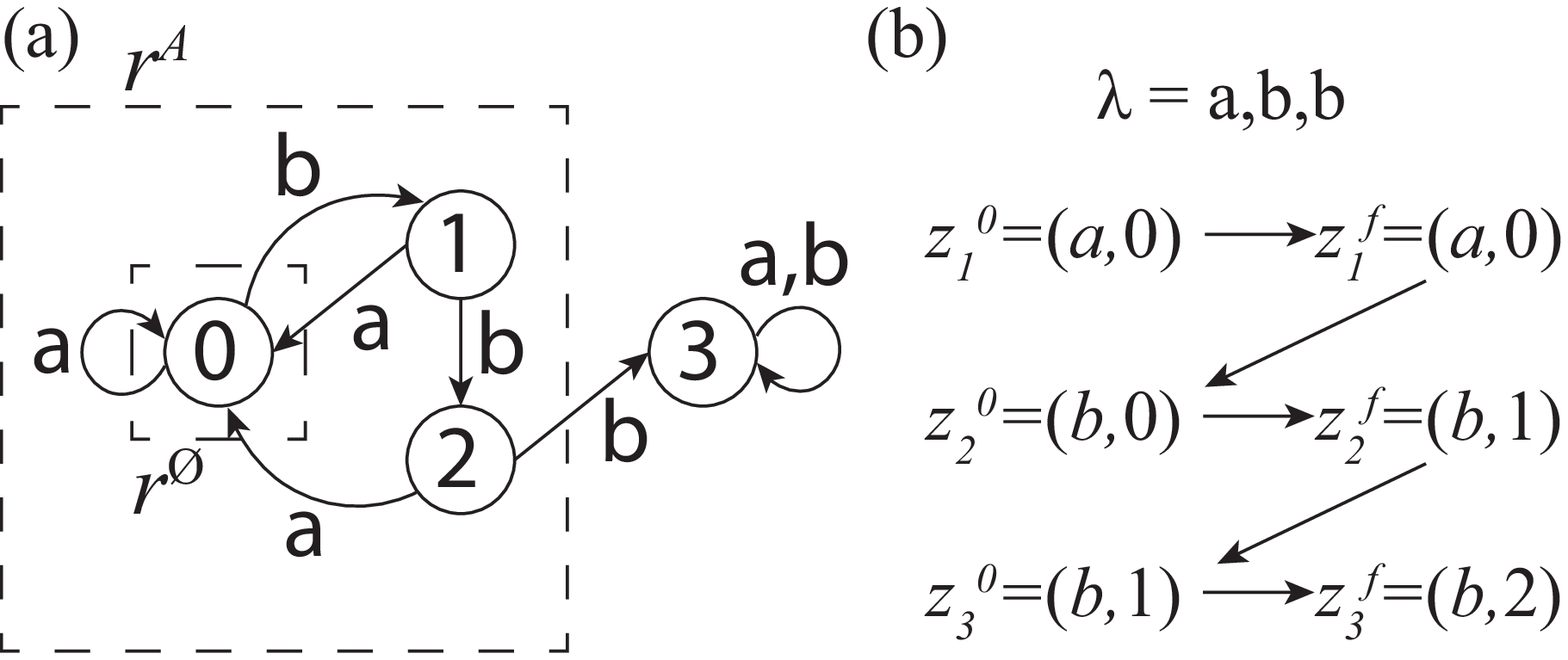}
    \caption{Example DFA with states $R=\{0,1,2,3\}$, alphabet $\Lambda=\{a,b\}$, initial state $r^\varnothing=0$ and accepting set $r^A=\{0,1,2\}$. The update function $\rho$ is illustrated in (a); the current computational state and the current input symbol specify the next computational state. This DFA accepts input strings that do not contain three or more consecutive $b$s. (b) shows the evolution of the local state through three iterations; the input string is read from left to right.}
    \label{fig:example_DFA}
\end{figure}

A DFA \cite{Hopcroft1979, Wolpert2019stoch, Moore2019} is  a 5-tuple  $(R, \Lambda, r^\varnothing, r^A, \rho)$ where: $R$ is a finite set of (computational) \textit{states}; $\Lambda$ is a finite \textit{alphabet} of input symbols; $\rho$ is a deterministic \textit{update function} specifying how the current DFA state is 
updated to a new one based on the next input symbol, i.e., $\rho: R \times \Lambda \rightarrow R$; $r^\varnothing \in R$ is a unique initial state; and $r^A \subset R$ is a set of \textit{accepting states}. An example is shown in Fig.~\ref{fig:example_DFA}.
The set of all finite input strings is indicated as  $\Lambda^*$. 

The DFA
 starts in state $r^\varnothing$ and an \textit{input string} $\lambda \in \Lambda^*$ is selected.
The selected input string's first symbol, ${\lambda}_1$, is then used to change the DFA's state to
$\rho({\lambda}_1, r^\varnothing)$. The computation
proceeds iteratively, with each successive component of
the vector ${\lambda}$ used as input to $\rho$ alongside the then-current DFA state to 
produce the next state. We write $\lambda_{-i}$ for the
entire vector $\lambda$ except for the $i$'th component.

We write
the DFA's computational state just before iteration $i$ as ${r}_{i-1}$, and we use ${r}_{i}$ for the state after the update.
The update in iteration $i$ is then the map
\eq{
({\lambda}_i, r_{i-1}) \rightarrow ({\lambda}_i,r_i) = \rho({\lambda}_i, r_{i-1}).
}
% during update $i$. 
We refer to this map as the \textit{local dynamics}, and define the set of
\textit{local states} as
\eq{
Z := R \times \Lambda,
}
with elements $z \in Z$. $z^0_i$ is the local state just before update $i$:
${z}^0_i = ({\lambda}_i,{r}_{i-1})$, and ${z}^f_i = ({\lambda}_i,{r}_{i})$ is the local state after update $i$. Note that ${z}^f_{i} \neq {z}^0_{i+1}$ in general, since 
${z}^0_{i+1}$ involves ${\lambda}_{i+1}$, not ${\lambda}_i$. The local update function fixes the full update function of the entire state space, since $\lambda_{-i}$ is unchanged during an update.

A DFA \textit{accepts} ${\lambda}$ if its state  is contained in $r^A$ after processing the final symbol.
The \textit{language} accepted by a DFA is the set of all input strings it accepts.
Many DFA accept the same language $L$; the \textit{minimal} DFA for $L$ has the smallest set of computational states $R$ for all DFA that accept $L$ \cite{Hopcroft1979,Moore2019}.

Fig.~\ref{fig:example_DFA}\,(a) shows a DFA with four computational states that processes words built from a two-symbol alphabet $\{a,b\}$. This DFA accepts all strings without three or more consecutive $b$s. Three iterations of this DFA when fed with an input $(a, b, b)$ are shown in Fig.~\ref{fig:example_DFA}\,(b). 
%The local state is updated three times by $\rho$; in between each of these updates, the next input symbol replaces the previous one.

DFA can be divided into those with an invertible local map $\rho$, and those with a non-invertible $\rho$. 
%$\rho$ is a deterministic special case of the conditional distribution $P$ introduced in Section~\ref{sec:stoch_thermo}. We can therefore use 
The map $\rho$ defines islands in the local state space:
an \textit{island} of  $\rho$ is a set of all inputs to $\rho$ that map to the same output (i.e.,
it the pre-image of an output of $\rho$). 
If the local dynamics defined by
$\rho$ is invertible, all local states are islands of size $1$; otherwise  $Z$ is partitioned by $\rho^{-1}$ into non-intersecting islands, some of which contain multiple elements. 
We write $c_i$ for the island that contains $z^0_i$. 
The DFA in Fig.~\ref{fig:example_DFA} is non-invertible, since $z^f_i=(a,0)$ could have arisen from either $z^0_{i}=(a,0)$, $(a,1)$ or $(a,2)$, which comprise an island.

\subsection{Thermodynamic description of DFA}
\label{sec:basic_thermo}

Details of our thermodynamic modelling of DFA are given in the Methods. In short, we assume that the logical states of the device are instantiated as well-defined, discrete physical states. At each iteration, a control protocol $\mu(t)$ is applied that drives a deterministic update of the DFA's state according to the logical rules of the computation. 

Although the overall update is deterministic, we assume that the input word is sampled from a distribution $p(\lambda)$, representing the possible computations that the DFA may be required to perform. We use $\boldsymbol{\lambda}$ to represent the random variable corresponding to the input word. The randomness of $\boldsymbol{\lambda}$ means that the computational state after update $i$; the local state before and after update $i$; and the island occupied during iteration $i$ are also random variables. To represent these random variables we use $\mathbf{r}_i$, $\mathbf{z}^0_i$, $\mathbf{z}^f_i$ and $\mathbf{c_i}$, respectively.

As outlined in the Methods, when a time-dependent control protocol $\mu(t)$ is applied to a thermodynamic system $X$ with a
finite set of states $\mathcal{X}=\{x_1,x_2,...\}$, the mismatch cost \cite{Wolpert2020,Kolchinsky2021,Wolpert2019stoch}
\begin{equation}
    \sigma_{\mu}(p) = D(p \,||\, q_\mu) - D(Pp\,||\,Pq_\mu)
    \label{eq:entropy_basic_main}
\end{equation}
is a lower bound on EP. Here, the time-dependent protocol $\mu(t)$ drives an evolution from $p(x)$ to $
p^\prime(x^\prime) = \sum_x P(x^\prime | x) p(x)$, or $p^\prime = Pp$. The 
distribution $q_\mu$ is known as the \textit{prior}  distribution~\cite{wolpert2016free,Wolpert2019stoch,Kolchinsky2020}, and is specific to the applied protocol $\mu(t)$.  $ D(p \,||\, q_\mu) $ is 
the Kullback-Leibler (KL) divergence 
between $p$ and
$q_\mu$; the mismatch cost is then the drop in KL divergence due to the matrix $P$. $\sigma_\mu$ is zero if $p = q_\mu$, and non-negative by the data processing inequality. Intuitively, the mismatch cost is the contribution to the EP of the  misalignment between  
the actual input distribution $p(x)$
and an optimal distribution $q_\mu(x)$ specified by the physical process $\mu(t)$. If the input distribution is well-matched to the protocol applied, $p(x)=q_\mu(x)$, EP is minimised. 

In the Methods, we outline how the EP of two co-evolving subsystems $X_a$ and $X_b$ that are not physically coupled during the period of evolution can be split into EP for the two subsystems in isolation, and a term related to the change in mutual information between the two. In the special case where $X_a$ evolves during the time period in question and $X_b=X_{-a}$ is static, the dynamics of ${X}_a$ under $\mu(t)=\mu_a(t)$ is said to be \textit{solitary} ~\cite{Wolpert2019stoch,Wolpert2020,Wolpert2020uncertainty}.
In this case, the mismatch cost is \cite{Wolpert2019stoch,Wolpert2020,Wolpert2020uncertainty,Boyd2018}
\eq{
\sigma &= \sigma_{\mu_a}(p_a)  - \Delta I.
% \\
  % &:= \sigma_{\rm mar}+\sigma_{\rm mod}
\label{eq:4a}
}  
Here, $p_a(x_a)$ is the initial marginal distribution for subsystem $a$, and $\Delta I$ is the change in the mutual information between $X_a$ and $X_{-a}$ over the period in question.

The first term in \cref{eq:4a} is the non-negative mismatch cost generated by ${X}_a$ running in
isolation, having marginalised over the other degrees of freedom. We call this the \textit{marginal} mismatch cost, $\sigma_{\rm mar}$. Like any other
mismatch cost, it is non-negative. The second 
term is the reduction in 
mutual information between ${X}_a$ and $X_{-a}$  \cite{Wolpert2019,Wolpert2020uncertainty},
which we call the \textit{modularity} mismatch cost, $\sigma_{\rm mod}$, after Ref. \cite{Boyd2018}. 
By the data processing inequality \cite{Elements_of_Information_Theory}, $\sigma_{\rm mod} \ge 0$.
Intuitively, this term reflects the fact that information about
the statistical coupling between ${X}_a$ and $X_{-a}$ is a store of non-equilibrium free energy, and that information is reduced in a solitary process.

To analyse the minimal thermodynamic costs of operating DFA under local and periodic constraints, we consider the effect of these constraints on the overall mismatch cost at each iteration. As discussed in the Methods, any additional entropy production can, in principle, be taken to zero. 

\subsection{Implementation of physical constraints}
\subsubsection{Locality}
In principle, one could build 
a DFA that physically couples the entire input word, ${\lambda}$, to the local subsystem ${z}_i$ during update $i$. However, this coupling is not required by the computational logic, which is local to ${z}_i$. Moreover, it would be extremely challenging to implement in practice; modern computers do not physically couple bits that do not need to be coupled by the logical operation in question. Accordingly, we assume that the evolution of the local state ${z}_i$ is solitary. As a result, the global mismatch cost splits into two non-negative components: a marginal mismatch cost associated with the evolution of the local state in isolation; and a modularity mismatch cost associated with non-conserved information between the local state and the rest of the system.

\subsubsection{Periodicity}
The marginal mismatch cost for iteration $i$ will depend on the similarity of $p(z_i^0)$, the initial distribution over local states, and $q_{\mu_i}(z^i_0)$, the prior distribution for the protocol $\mu_i(t)$ implemented at iteration $i$. Typically, $p(z_i^0)$ will vary with $i$. In theory, one could design $\mu_i(t)$ to match these variations, ensuring $q_{\mu_i}(z^i_0)=p(z_i^0)$ at each update, eliminating $\sigma_{\rm mar}$.  However, designing such a protocol would require knowledge of $p(z_i^0)$ --- which in turn would require running a computation emulating the DFA before running the DFA, gaining nothing. Moreover, one of the major strengths of computing paradigms such as DFA, Turing machines and real world digital computers is that their logical updates are not iteration-dependent.
It is therefore natural to impose a second constraint: the protocol $\mu_i(t)$, like the logical update $\rho$, is identical at each update $i$ ($\mu_i(t) = \mu(t)$). 
% , i.e., it is required to be ``cyclic''. 
Formally, we define a \textit{local, periodic} DFA (LPDFA) as any process that implements a DFA via a repetitive, solitary process on the local state ${z}_i$.

\subsection{General consequences of local and periodic constraints}
We briefly consider the consequences of locality and periodicity in general, before re-focussing on DFA. The mismatch and modularity costs introduced in Section~\ref{sec:basic_thermo} are well established. However, systems that perform non-trivial computations by iterating simpler logical steps on subsystems are exposed to these costs in a way that simpler operations, like erasing a bit, are not. The need to operate iteratively on an input that is evolving from iteration to iteration makes the mismatch cost unavoidable. Additionally, modularity-cost-inducing statistical correlations result from the need to carry information between iterations, which will not be required in simpler systems.  

Consider a physical realisation of an arbitrary computation that is local and periodic in a way that reflects the locality and periodicity of the computational logic. Then the marginal and modularity mismatch costs set a lower bound on EP, regardless of any further details about how the computation is implemented. Specifically, let $X$ be the computational system and  ${X}_i$ the local subsystem that is updated at iteration $i$. Then over the course of $N$ iterations, the system will experience a total marginal mismatch cost
\eq{
\sigma_{\rm mar}=\sum_{i=1}^{N}
       D(p(x_i) \,||\, q_\mu(x_i)) - D(Pp(x_i)\,||\,Pq_\mu(x_i)),
\label{eq:sum_mismatch_costs}
}
where $P$ is the update matrix, $p(x_i)$ is the initial distribution of the local state and $q_\mu(x_i)$ is the prior built in to the actual protocol $\mu(t)$.

Eq.~\ref{eq:sum_mismatch_costs} depends on the details of $\mu(t)$ beyond the locality and periodicity constraints. However, some choice of $q_\mu$ (and hence $\mu(t)$) will minimize $\sigma_{\rm mar}$, setting a lower bound on EP that is independent of these details.
\eq{
\sigma_{\rm mar} &\geq \underset{q_\mu}{\min} \sum_{i=1}^{N}
       D(p(x_i) \,||\, q_\mu(x_i)) - D(Pp(x_i)\,||\,Pq_\mu(x_i)) \nonumber  \\
       & \geq 0.
\label{eq:sum_mismatch_costs2}
}
Unless $p(x_i)$ is identical for all $i$, or $P$ is a simple permutation, it is not generally possible to choose a single $q_\mu$ that will eliminate $\sigma_{\rm mar}$ at every iteration $i$. In this case, Eq.~\ref{eq:sum_mismatch_costs2} provides a strictly positive periodicity-induced lower bound on the EP that depends purely on the logic of the computation performed.

Similarly, the accumulated modularity cost follows directly as 
\begin{equation}
    \sigma_{\rm mod} = -\sum_{i=1}^N \Delta I(X_i ;X_{-i}) \geq 0,
    \label{eq:sum_mod_costs}
\end{equation}
where $\Delta I(X_i ;X_{-i})$ is the change in mutual information between ${X}_i$ and ${X}_{-i}$ due to update $i$. As with Eq.~\ref{eq:sum_mismatch_costs2}, this contribution to EP is entirely determined by the  computational paradigm used and the distribution of inputs; it is independent of the details of the implementation given the assumption of locality and periodicity. Taken together, the sum of $\sigma_{\rm mar}$ and $\sigma_{\rm mod}$ from Eq.~\ref{eq:sum_mismatch_costs2} and Eq.~\ref{eq:sum_mod_costs} constitute a strengthened second law for periodic, local computations that depends only on the logic of the computation, not the details of its implementation.

These implementation-independent lower bounds, alongside the qualitative observation that computing systems are particularly vulnerable to modularity and mismatch costs, is the first main result of this work.
These results apply to any computational system implemented using a periodic, local process. For
the rest of the paper, we will focus on DFA. Doing so allows us to illustrate the consequences of local and periodic restrictions in a concrete computational model.

\begin{figure*}
    \centering
    \centering
    \begin{subfigure}{0.48\textwidth}
        \includegraphics[width=\textwidth]{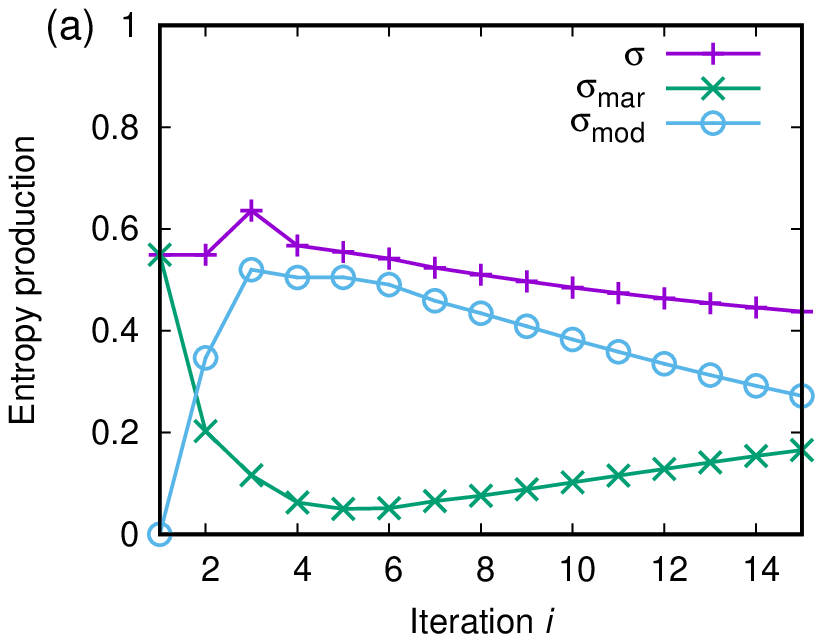}
    \end{subfigure}
\hfill
\begin{subfigure}{0.48\textwidth}
    \includegraphics[width=\textwidth]{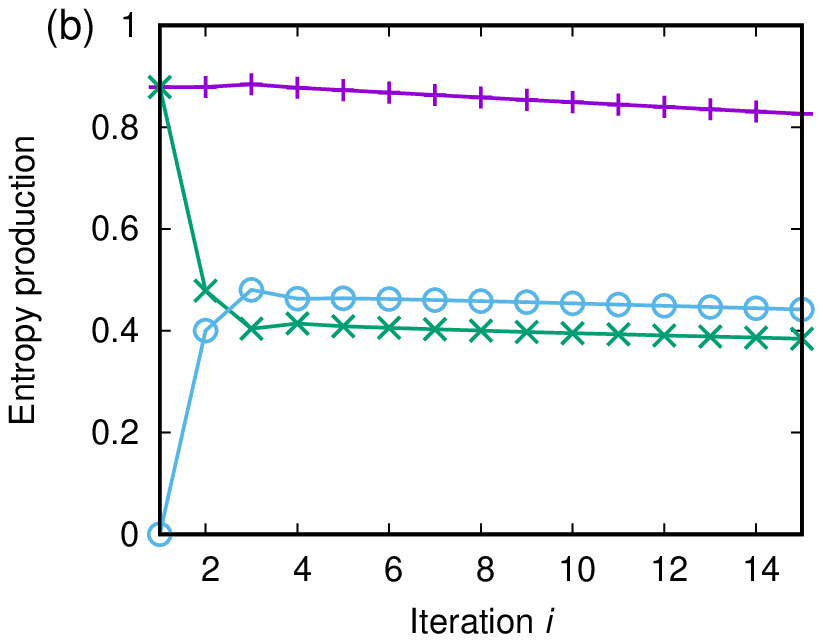}
\end{subfigure}
\hfill
\begin{subfigure}{0.48\textwidth}
    \includegraphics[width=\textwidth]{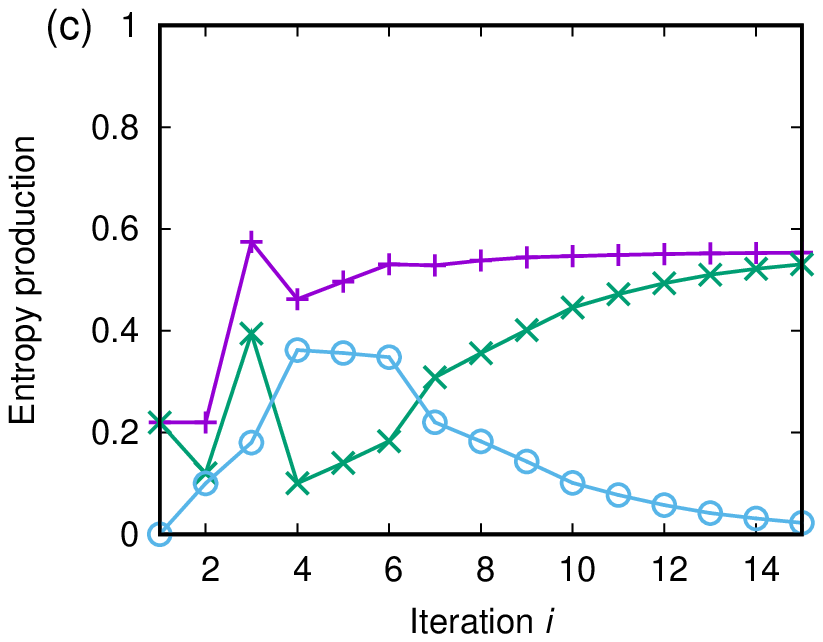}
\end{subfigure}
\hfill
\begin{subfigure}{0.48\textwidth}
    \includegraphics[width=\textwidth]{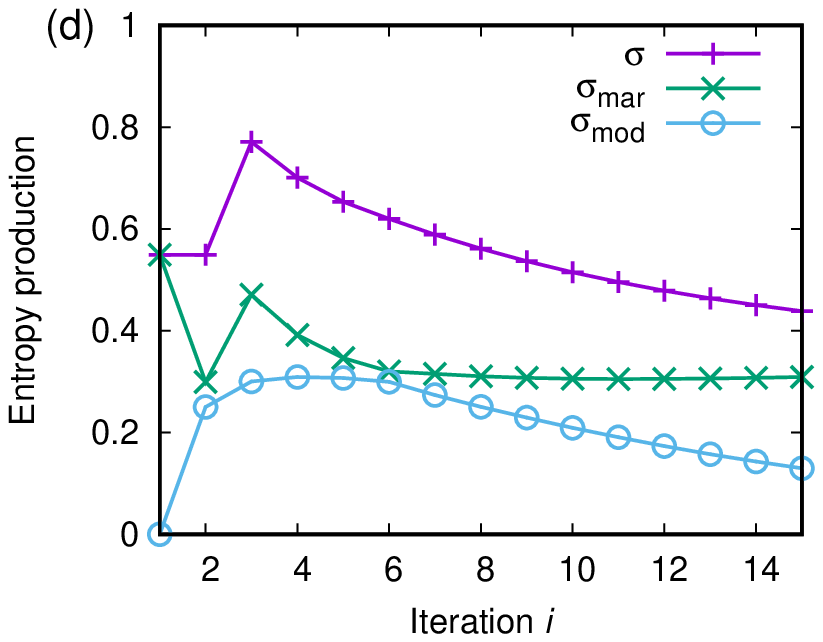}
\end{subfigure}
\caption{FIX LABELS
EP in a simple system shows non-trivial dependence on iteration and input word distribution. We plot total EP $\sigma^i$, and its decomposition into $\sigma_{\rm mar}^i$ and $\sigma^i_{\rm mod}$, for the DFA in Fig.~\ref{fig:example_DFA}\,(a), which accepts all words that do not contain three or more consecutive $b$s. In all cases we use a uniform prior $q_\mu(z_i^0|c_i)$ within each island, and consider a distribution of input words with fixed length $N=15$, but vary the distribution of input words $p(\lambda)$. (a) input words have independent and identically distributed (IID) symbols with $p(a)=p(b) =0.5$. (b) input words have IID symbols with $p(a) = 0.8$ and $p(b) = 0.2$. (c) input words have IID symbols with $p(a)=0.2$ and $p(b)=0.8$. (d) input words are Markov chains. The first symbol is $a$ or $b$ with equal probability, and subsequently $P(\boldsymbol{\lambda_{i+1}} = \boldsymbol{\lambda_i})$ = 0.8.}
    \label{fig:examp_ent_prodn}
\end{figure*}

\subsection{Entropy production for LPDFA}
Under our assumptions, the EP when applying a solitary dynamics $\mu(t)$ to an initial distribution $p(z_i^0, {\lambda}_{-i})$ at the update stage of iteration $i$ of a DFA is 
% \cite{Wolpert2020uncertainty}
\begin{equation}
\begin{array}{c}
\sigma^i_{{\mu}}(p(z_i^0, {\lambda}_{-i}))
% \sigma_{\rho,q_{\mu}}(p(z_i^0, {\lambda}_{-i} ))
= \sigma_{\rm mar}^i + \sigma_{\rm mod}^i, 
    % \sigma_{\mu_i}(p(z_i^0, {\lambda}_{-i}))=
    % D(p(z_i^0) \,||\, q_{\mu_i}(z_i^0) ) - D(p(z_i^f)\,||\,q_{\mu_i}(z_i^f)) \\
    % \\
    % % +  \sum_{c_i} p(c_i) \sigma_{\mu_i}(q_{\mu_i}^{c_i}) 
    % +  I(Z^0_i; {\Lambda}_{-i}) - I(Z^{f}_i; {\Lambda}_{-i}).
    \end{array}
        \label{eq:EP1}
\end{equation}
where 
\eq{
\sigma_{\rm mar}^i := D(p(z_i^0) \,||\, q_{\mu}(z_i^0) ) - D(p(z_i^f)\,||\,q_{\mu}(z_i^f))
 \label{eq:EP_local}
}
is the marginal mismatch cost of update $i$,
and 
\eq{
\sigma_{\rm mod}^i := I(\mathbf{z}^0_i; \boldsymbol{\lambda}_{-i}) - I(\mathbf{z}^{f}_i; \boldsymbol{\lambda}_{-i})
\label{eq:EP_mod}
}
is the modularity mismatch cost of update $i$. 
%At first glance it might appear that
%\cref{eq:EP1} contradicts Eq.\,6 in~\cite{Wolpert2020}. It
%is shown in the Methods that in fact they are equivalent.
A variant of the modularity cost in \cref{eq:EP_mod} 
% mismatch cost
was considered in isolation in Ref.~\cite{Ganesh2013}, for 
the special case of DFA operating in steady state. 

Henceforth, for simplicity, we suppress the dependence of $\sigma^i$ on $\mu$ since $\mu$ is constant over all iterations.
The KL divergences in Eq.~\ref{eq:EP_local}, giving $\sigma_{\rm mar}$, can be simplified for LPDFA. 
Since each update in an LPDFA deterministically collapses all probability within an island to one state, $p(z^f_i|c_i) = q(z^f_i|c_i)$. As shown in Section 2
I of the Supplementary Information, this simplification implies that
\begin{align}
 \sigma_{\rm mar}^i =
    & \sum_{c_i} p(c_i) D ( p(z^0_i|c_i)\,||\,q_\mu(z^0_i|c_i) ),
    \label{eq:KLD2}
\end{align}
which is the second main result of this work. $ \sigma_{\rm mar}^i$ is therefore the divergence between initial  and prior distributions, conditioned on the island of the initial state. 

In Fig.~\ref{fig:examp_ent_prodn}, we explore the properties of $\sigma_{\rm mar}^i$ for the DFA shown in Fig.~\ref{fig:example_DFA}. The four sub-figures show $ \sigma_{\rm mar}^i$ for four distinct distributions $p(\lambda)$, and a fixed (uniform) prior $q_\mu$. We immediately see that $ \sigma_{\rm mar}^i$ is strongly dependent on both the distribution of input words and the iteration, with $ \sigma_{\rm mar}^i$  non-monotonic in $i$ in all four cases.

$\sigma_{\rm mar}^i$  is determined by a combination of how well tuned the prior is to the input distribution  within a given island, and the probability of that island at each iteration. At the start of iteration 1, particularly for subfigure (b), there is a high probability of the system being in the island $\{(a,0); (a,1); (a,2)\}$ and the uniform prior is poorly aligned with the actual initial condition within this island (all in $(a,0)$). At larger $i$, this cost drops both because the probability of being in that island drops, and the conditional distribution within the island gets more uniform. 

For iterations $i \geq 3$, the system has a non-zero probability of being in the other non-trivial island $\{(b,2); (b,3)\}$. The uniform prior is initially poorly matched to the conditional distribution within this island (at the start of iteration $i=3$, the system cannot be in $(b,3)$). Additionally, the probability of the system being in this island is quite low for subfigure (a) and (b), but much higher for (c) and (d) -- explaining the jumps in those traces.

The third main result of this work is a simple expression for the modularity mismatch cost for DFA. As we show  in Section 3 of the Supplementary Information,
\eq{
    \sigma_{\rm mod}^i = H(\mathbf{z}^0_i|\mathbf{c}_i).
    \label{eq:4}
}
Surprisingly, $\sigma_{\rm mod}^i$, a \textit{global} quantity, is given by the 
% reduction in 
entropy of the \textit{local} state at the beginning of the update, conditioned on
the island occupied at the start of iteration $i$. This result holds regardless of the distribution of input strings or the DFA's complexity. 

To understand  Eq.~\ref{eq:4} intuitively, we note that $\mathbf{z}^0_i$ in general contains information about
$\boldsymbol{\lambda}_{-i}$. After the update, any information provided by $\boldsymbol{\lambda}_i$ alone is retained, since the input symbol is not updated by the DFA. Moreover, for islands of size 1, the combined values of $\boldsymbol{\lambda}_i$ and $\mathbf{r}_i$ are just as informative about $\boldsymbol{\lambda}_{-i}$ as $\boldsymbol{\lambda}_i$ and $\mathbf{r}_{i-1}$ were. However, for non-trivial islands, the extra information provided by $\mathbf{r}_{i-1}$ on top of $\boldsymbol{\lambda}_i$  is lost, yielding  Eq.~\ref{eq:4}. 

%The precise form of Eq.~\ref{eq:4} relies upon the fact that the current state of the DFA can be predicted exactly from the current symbols of the input word. We would not, therefore, expect the precise form to apply directly to more complex systems such as TMs, which can overwrite their input program. 

We see from our example system in Fig.~\ref{fig:examp_ent_prodn} that modularity costs behave very differently from marginal mismatch costs. In general, $\sigma^i_{\rm mod}$ tends to zero as the probability of being absorbed into state 3 increases: in this case, there is no entropy of $\mathbf{z}_i^0$. Modularity costs stay high for system (b), in which $bbb$ substrings are infrequent.

\begin{figure}[!]
    \centering
    \centering
    \begin{subfigure}{0.3\textwidth}
        \includegraphics[width=\textwidth]{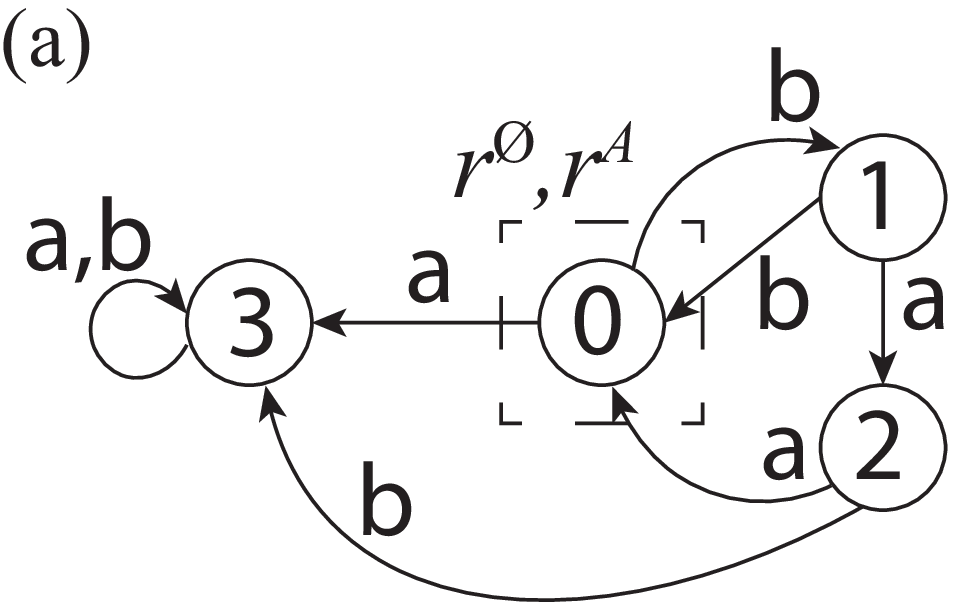}
    \end{subfigure}
\hfill
\begin{subfigure}{0.35\textwidth}
    \includegraphics[width=\textwidth]{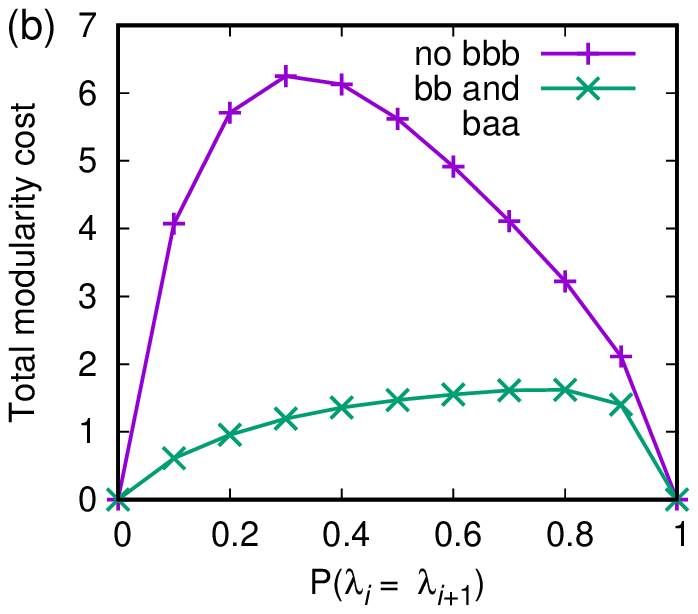}
\end{subfigure}
\hfill
\begin{subfigure}{0.35\textwidth}
    \includegraphics[width=\textwidth]{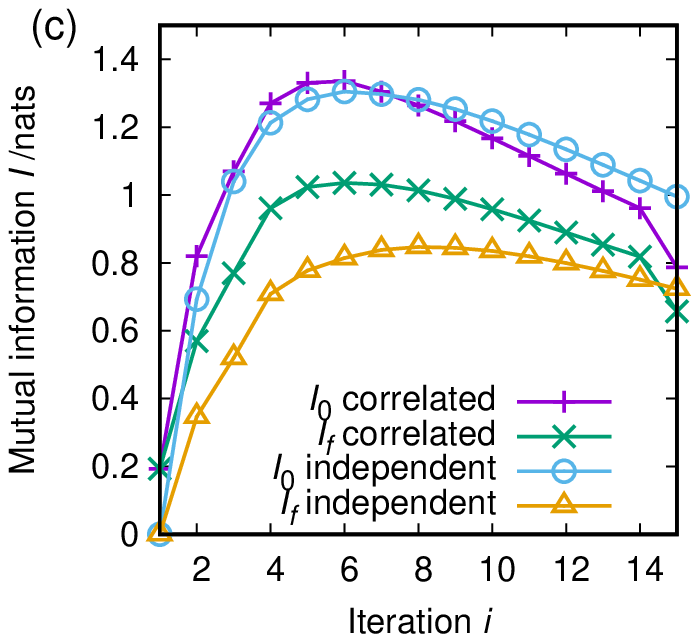}
\end{subfigure}
\caption{Correlated input words do not generate high modularity costs. (a) A 4-state DFA that processes words formed from a two-symbol alphabet, accepting those formed by concatenating $bb$ and $baa$ substrings. (b) Total modularity cost $\sum_{i=1}^N \sigma_{\rm mod}^i$ for the DFA in (a) and the DFA in Fig.~\ref{fig:example_DFA}\,(a), when processing words of length $N=15$ that are generated using a Markov chain. Modularity cost is plotted as a function of the probability that subsequent symbols in the word have the same value. (c)  Mutual information between the local state and the rest of the input word before ($I_0$) and after ($I_f$) the update of iteration $i$, for the DFA in Fig.~\ref{fig:example_DFA}\,(a). Data is plotted for $P(\boldsymbol{\lambda_{i+1}} = \boldsymbol{\lambda_i})=0.8$ (correlated) and $P(\boldsymbol{\lambda_{i+1}} = \boldsymbol{\lambda_i})=0.5$ (independent). }
    \label{fig:effect_of_correlations}
\end{figure}

Modularity costs are relatively low in Fig.~\ref{fig:examp_ent_prodn} (d), in which symbols of the input word are correlated. Na\"{i}vely, one might have assumed that a larger $I(\mathbf{z}_i^0, \boldsymbol{\lambda}_{-i})$ generated by a correlated input word would be more susceptible to large modularity costs.  We explore this question in more detail in Fig.~\ref{fig:effect_of_correlations} for both the DFA illustrated in Fig.~\ref{fig:example_DFA}\,(a) and a second DFA that accepts words that are concatenations of $bb$ and $baa$ substrings (Fig.~\ref{fig:effect_of_correlations}\,(a)).

In Fig.~\ref{fig:effect_of_correlations}\,(b) we plot the total modularity cost, $\sum_{i=1}^N \sigma_{\rm mod}^i$, for both DFA processing a Markovian input, as a function of the degree of correlation, $P(\boldsymbol{\lambda_{i+1}} = \boldsymbol{\lambda_i})$. We see that in both cases, the uncorrelated input words with $P(\boldsymbol{\lambda_{i+1}} =  \boldsymbol{\lambda_i})=0.5$ have relatively high (though not maximal) modularity cost, and fully correlated strings have $\sigma_{\rm mod}=0$. 

To understand why, consider Fig.~\ref{fig:effect_of_correlations}\,(c), in which we plot the information between the local state and the rest of the input word before ($I_0= I(\mathbf{z}_i^0;\boldsymbol{\lambda}_{-i})$) and after ($I_f= I(\mathbf{z}_i^f;\boldsymbol{\lambda}_{-i})$) the update of iteration $i$, for the original DFA in Fig.~\ref{fig:example_DFA}\,(a). We consider uncorrelated input words ($P(\boldsymbol{\lambda_{i+1}} = \boldsymbol{\lambda_i})=0.5$) and moderately correlated input words ($P(\boldsymbol{\lambda_{i+1}} = \boldsymbol{\lambda_i})=0.8$). At early iterations, $I_0$ is larger for the correlated input, as would be expected (at later times, the DFA with correlated input is more likely to be absorbed into state 3, reducing $I_0$). More importantly, the system with correlated inputs retains more of its information in the final state. Because $\boldsymbol{\lambda}_{-i}$ is correlated with the the current symbol $\boldsymbol{\lambda}_i$, it is a better predictor of the final state of the update. In the limit of $P(\boldsymbol{\lambda_{i+1}} = \boldsymbol{\lambda_i})=1$ or 0, there is no modularity cost as $\mathbf{z}_i^f$ is perfectly predictable from $\boldsymbol{\lambda}_{-i}$.

Combining  \cref{eq:KLD2,eq:4}
% and $\sigma_{\rm mod} = H(Z^0_i|C_i)$ can be combined to yield
gives
\begin{equation}
    \sigma^i(p(z_i^0, {\lambda}_{-i}))  = \sum_{c_i} p(c_i) H(p(z_i^0|c_i)\,||\,q_\mu(z_i^0|c_i)),
    \label{eq:cross}
\end{equation}
where
\eq{
H(p(z_i^0|c_i)\,||\,q_\mu(z_i^0|c_i))
    := -\sum_{z_i^0 \in c_i} p(z_i^0|c_i) \ln q_\mu(z_i^0|c_i)
    }
is the \textit{cross entropy} between $q_\mu(z_i^0|c_i)$ and $p(z_i^0|c_i)$.  This total entropy production is also shown for the example DFA of Fig.~\ref{fig:example_DFA}\,(a) in Fig.~\ref{fig:examp_ent_prodn}.

\subsection{Reducing the marginal mismatch cost through choice of priors}

\begin{figure}[!]
    \centering
    \centering
    \begin{subfigure}{0.4\textwidth}
        \includegraphics[width=\textwidth]{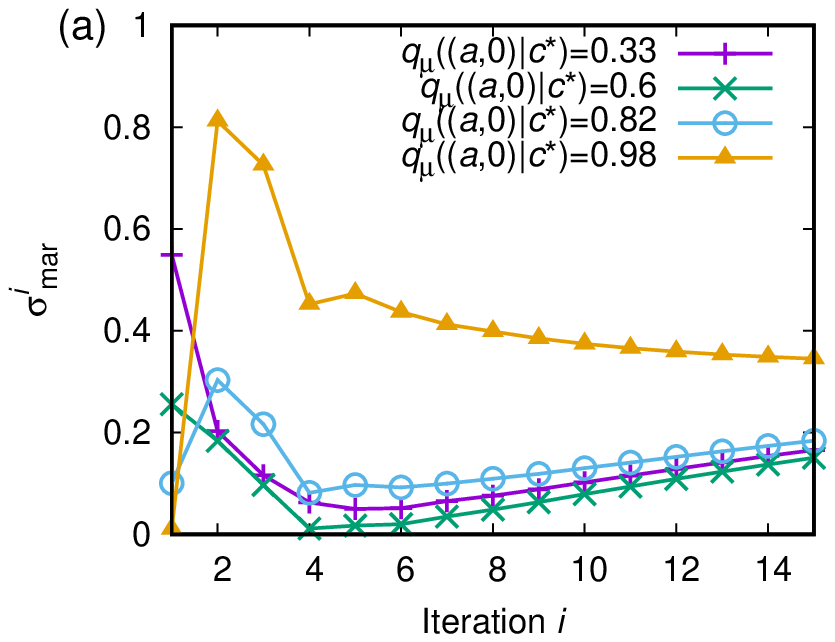}
    \end{subfigure}
\hfill
\begin{subfigure}{0.4\textwidth}
    \includegraphics[width=\textwidth]{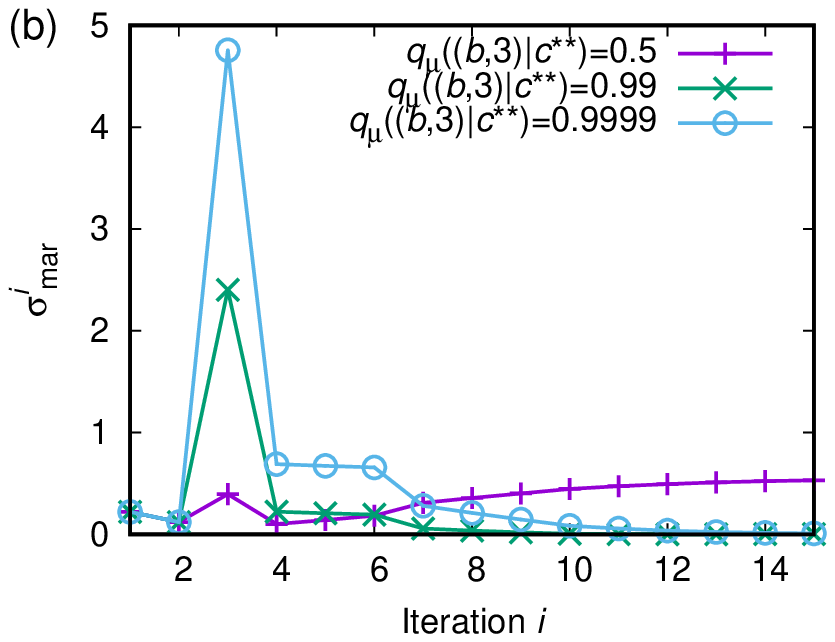}
\end{subfigure}
\caption{ Applying a biased prior $q_\mu(z^0_i|c_i)$ can reduce the local mismatch cost. (a) $\sigma_{\rm mar}^i$ for the DFA in Fig.~\ref{fig:example_DFA}\,(a), given input words of length $N=15$ with IID symbols ($p(a)=p(b) =0.5$). Results are plotted for different values of $q_\mu((a,0)|c^\star)$, where $c^\star= \{(a,0),(a,1),(a,2)\}$ is the island containing $(a,0)$. $q_\mu(z^0_i|c_i)$ is otherwise unbiased, and $q_\mu((a,0)|c^\star)=1/3$ corresponds to a totally unbiased prior. (c) Equivalent to (a), but for input $p(a)=0.2$, $p(b) =0.8$, and applying a bias to $q_\mu((b,3)|c^{\star \star})$, where $c^{\star \star}= \{(b,2),(b,3)\}$ is the other non-trivial island for this DFA. $q_\mu(z^0_i|c_i)$ is otherwise unbiased, and $q_\mu((b,3)|c^{\star \star})=1/2$ corresponds to a totally unbiased prior.
}
    \label{fig:priors}
\end{figure}

\subsubsection{Applying a bias to the prior}
\label{sec:bias}

It is natural to ask how $q_\mu(z^0_i|c_i)$ might be chosen to minimize EP for a given $p(\lambda)$ and a
given DFA. One might hope that $q_\mu(z^0_i|c_i)$ could be tuned to  $p(\lambda)$ alone, without any reference to the operation of the DFA. Unfortunately, however, such an approach will fail. The states within each island all have the same value of $\lambda$, because the update map $\rho(\lambda_i, r_{i-1}) = ({\lambda}_i, r_{i-1}) \rightarrow ({\lambda}_i,r_i)$ does not update the input symbol. Applying a prior that is a function of $\lambda$ alone results in a uniform $q_\mu(z^0_i|c_i)$.

Reducing the mismatch cost through choice of prior thus requires some understanding of the computational state, not just the inputs. For example, for the DFA in Fig.~\ref{fig:example_DFA}\,(a), the computation starts in the state $r^\varnothing=0$. Biasing $q_\mu(z^0_i|c_i)$ towards states with $r=0$, as we show in Fig.~\ref{fig:priors}\,(a), can reduce the marginal mismatch cost of the first step. If the bias is too strong, then increased costs at later iterations overwhelm the initial reduction. It is possible, however, to reduce the total EP with a moderate bias of $q_\mu(z^0_i|c_i)$  towards states with $r=0$.

Alternatively, one could bias $q_\mu(z^0_i|c_i)$ towards states with $r=3$, since most trajectories will eventually be absorbed. As shown in Fig.~\ref{fig:priors}\,(b), doing so incurs an extra cost at short times, particularly at iteration $i=3$. At the start of the third iteration, the DFA is moderately likely to be in computational state $r=2$, but cannot be in computational state $r=3$, so the biased prior is a poor match for  $p(z^0_i|c_i)$. At later iterations, however, the biased prior performs better.
Again, a moderate  bias performs best overall.

\subsubsection{Advantages of a uniform prior}
\label{sec:uniform}

Section~\ref{sec:bias} shows that it is possible to reduce EP by applying biased priors. However, we also saw that very biased priors could lead to very high  EP. As noted in Ref.~\onlinecite{Riechers2021impossibility},
in which a similar result to Eq.~\ref{eq:KLD2} was derived in the absence of distinct islands, $\sigma^i_{\rm mar}$ penalizes an over-confident prior $q_\mu(z^0_i|c_i)$. If $q_\mu(z^0_i|c_i)=0$ for a given state but $p(z^0_i|c_i)\neq0$, Eq.~\ref{eq:cross} implies $\sigma^i_{\rm mar} \rightarrow \infty$. The authors of Ref.~\onlinecite{Riechers2021impossibility} hypothesised, therefore, that a uniform $q_\mu(z^0_i|c_i)$ may be optimal.

As a fourth main result of this work, we present three  important properties of a $q_\mu(z_i^0|c_i)$ that is uniform for each $c_i$, {\it i.e.}, a prior
$q_\mu(z_i^0|c_i) = 1/L_{c_i}$, with $L_{c}$ the size of island $c$. First, for such a prior, Eq.~\ref{eq:cross} becomes
\begin{align}
 \sigma^i(p(z_i^0, {\lambda}_{-i}))  &=
    \langle \ln L_{c_i} \rangle \leq \ln L_{c_{\rm max}}.
    \label{eq:KLD3}
\end{align}
Here, $L_{c_{\rm max}}$ is the size of the largest island of $\rho$. Eq.~\ref{eq:KLD3} gives a  finite upper bound to EP for LPDFA employing a uniform prior distribution $q_\mu(z^0_i|c_i)=1/L_{c_i}$, constrained by the size of the largest island.

Second, for any protocol, the worst case EP is at least $\ln L_{c_{\rm max}}$. A uniform prior distribution $q_\mu(z^0_i|c_i)= 1/L_{c_i}$ therefore minimizes the worst case EP. To verify this claim, consider the input distribution $p(z_0^i)= \delta_{z_0^i, z_{\rm min}}$, where $z_{\rm min}$ is a state that minimizes $q_\mu(z^0_i|c_i)$ within the largest island. For such a distribution,  Eq.~\ref{eq:KLD3} reduces to
\begin{equation}
\sigma^i(p(z_i^0, {\lambda}_{-i})) = - \ln q_\mu(z_{\rm min}|c_{\rm max}) \geq \ln L_{c_{\rm max}},
\end{equation}
where the final inequality follows from $q_\mu(z_{\rm min}|c_{\rm max}) \leq 1/L_{c_{\rm max}}$.

Finally, the uniform prior distribution  $q_\mu(z^0_i|c_i)= 1/L_{c_i}$ minimizes predicted average EP if a designer is maximally uncertain about $p(z_i^0, {\lambda}_{i})$.
A designer may not know that $p_i(z^0_i|c_i)$ is the input distribution at iteration $i$  -- either because $p({\lambda})$, or the  DFA's dynamics on $p({\lambda})$, are unknown. Thus the choice of protocol $\mu(t)$, and hence $q_\mu(z_i^0|c_i)$, is performed under uncertainty over not just the input state, but also the distribution from which that state is drawn.  

Let the designer's belief about the distributions $p(c_i)$ and $p(z^0_i|c_i)$ be represented by a distribution $\pi(v,w)$ over an (arbitrary) discrete set of possible distributions indexed by $v,w$: $p_v(c_i)$, $p_w(z^0_i|c_i)$. The designer's best estimate of the expected EP at iteration $i$ is then (see Section 4 of the Supplementary Information)
 \begin{align}
\hat{\sigma}^i &
= H(\mathbf{z}_i^0|\mathbf{c}_i,\mathbf{v},\mathbf{w}) + I( \mathbf{z}_i^0;\mathbf{w}|\mathbf{c}_i,\mathbf{v}) \nonumber \\
& +\sum_v \pi(v) \sum_{c_i} p_v(c_i) D( \hat{p}(z_i^0 |c_i,v)  \,||\,  q_\mu(z_i^0 |c_i)).
\label{eq:KLD6}
\end{align}
Here, $H(\mathbf{z}_i^0|\mathbf{c}_i,\mathbf{v},\mathbf{w})$ and $I( \mathbf{z}_i^0;\mathbf{w}|\mathbf{c}_i,\mathbf{v})$ are defined with respect to the estimated joint distribution, $\hat{p}(v,w,z_i^0,c_i)= \pi(v)\pi(w|v)p_v(c_i)p_w(z_i^0 |c_i)$, and $\hat{p}(z_i^0 |c_i,v) = \sum_w \pi(w|v){p}_w(z_i^0 |c_i) $ is  the designer's estimate for the probability distribution within an island, having averaged over the uncertainty quantified by $\pi(w|v)$. 

All three terms in Eq.~\ref{eq:KLD6} are non-negative. The first is $\sigma^i_{\rm mod}$ averaged over $\mathbf{v}$ and $\mathbf{w}$. The third is the marginal mismatch cost between $\hat{p}(z_i^0 |c_i,v)$ and  $q_\mu(z_i^0 |c_i)$.  However, even if $q_\mu(z_i^0 |c_i)$ matches the average estimated distribution within an island, $\hat{p}(z_i^0 |c_i,v)=q_\mu(z_i^0 |c_i)$, the best estimate of $\hat{\sigma}_{\rm mar}^i$ is non-zero. The second term, $I( \mathbf{z}_i^0;\mathbf{w}|\mathbf{c}_i,\mathbf{v})$, quantifies how much uncertainty in $\mathbf{w}$ is actually manifest in an uncertainty in the input distribution; variability about $\hat{p}(z_i^0 |c_i,v)$ gives positive expected EP. An equivalent term was previously identified in Ref.~\cite{korbel2021} for arbitrary processes with a single island. 

$H(\mathbf{z}_i^0|\mathbf{c}_i,\mathbf{v},\mathbf{w})$ and $I( \mathbf{z}_i^0;\mathbf{w}|\mathbf{c}_i,\mathbf{v})$ are protocol-independent and cannot be changed for a given computation. $D( \hat{p}(z_i^0 |c_i,v)  \,||\,  q_\mu(z_i^0 |c_i))$, however, can be minimized by by choosing $q_\mu(z_i^0 |c_i)=\hat{p}(z_i^0 |c_i,v)$. Given maximal uncertainty, the designer's best estimate will be uniform: $\hat{p}(z_i^0 |c_i,v)=1/L_{c_i}$. In this case, a uniform $q_\mu(z_i^0 |c_i)=1/L_{c_i}$ minimizes estimated average EP.

 The results hitherto  apply to LPDFA, but do not reflect the actual computation performed. The results for $\sigma^i_{\rm mar}$ -- including the optimality of a uniform protocol -- apply to any deterministic process; the LPDFA's restrictions simply justify why $q_{\mu_i}(z_i^0 |c_i)$ cannot be tuned to $p(z^0_i|c_i)$ at each $i$. The results for $\sigma_{\rm mod}^i$ are more specific, relying on a solitary process using a single symbol $\lambda_i$ from an unchanging ``input string", and a device whose state after the update is unambiguously specified by $\lambda_{i\leq j}$. Nonetheless,
 $\sigma_{\rm mod}^i$ in Eq.~\ref{eq:4} is not directly related
 to the computational task. We now explore how EP is related to $\rho$, and the language accepted by the DFA.

\subsection{Relating EP to computational tasks}
\subsubsection{Nonzero EP is common in non-invertible LPDFA}
The EP in Eq.~\ref{eq:cross} is positive iff $q_\mu(z^0_i|c_i)\neq 1$ for any $z_i^0,c_i$ for which  $p(z^0_i|c_i) \neq 0$ and  $p(c_i) \neq 0$. This condition is met whenever an island $c_i$ 
with  $p(c_i) > 0$ has at least
two elements $z_0$ with $p(z^0_i|c_i) > 0$.
There are two ways to avoid this EP. One is if all islands have
a single element, {\it i.e.}, the local update function
$\rho$ is invertible (this observation was made for $\sigma_{\rm mod}$ alone in Ref.~\cite{Ganesh2013}).  The second is if the distribution of input strings $p({\lambda})$ is such that for
every island $c_i$ with at least two elements, all but one of those elements always have $p(z^0_i|c_i) =0$. However, in that case, $q_\mu(z^0_i|c_i)$ must be finely-tuned to match this condition when the physical system
implementing the computation is constructed. 
As discussed in Sections~\ref{sec:bias} and \ref{sec:uniform}, this  strategy risks high costs for overconfidence.

\begin{figure}
    \centering
    \includegraphics[width=8cm]{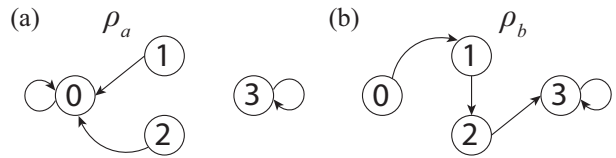}
    \caption{Decomposition of the DFA in Fig.~\ref{fig:example_DFA}\,(a) into networks of transitions for each input symbol, $\rho_\lambda$. (a) Network for $\lambda_i=a$, where the state $r=0$ is a conflict state. (b) Network for $\lambda_i=b$, where the state $r=3$ is a conflict state.}
    \label{fig:example_decomposed}
\end{figure}

We now focus on the former way of achieving zero EP, asking what determines whether $\rho$ is invertible.
Since $\rho$ preserves the input symbol ${\lambda}_i$, it can only be non-invertible if it maps two distinct  computational states
to the same output for the same symbol $\lambda_i$.
If we illustrate $\rho$  by a series of directed graphs, one for each value of $\lambda_i$, then a non-invertible DFA will have at least one state with at least two incoming transitions for at least one value of $\lambda_i$. 
We label states with more than one incoming transition for a given $\lambda_i$ as {\it conflict states}; conflict states for the DFA in Fig.~\ref{fig:example_DFA}\,(a) are shown in Fig.~\ref{fig:example_decomposed}.

\subsubsection{The minimal DFA for a given language does not generally minimize or maximize EP}
The minimal DFA for a language $L$ has the smallest set of computational states $R$ for all DFA that accept $L$.
This minimal DFA has just enough memory to sort parsed substrings into classes of equivalent strings, so that information can be passed forward to complete the computation \cite{Hopcroft1979,Moore2019,nerode1958}.
More formally, define input strings ${\lambda}$ and ${\mu}$ to be \textit{equivalent}
with respect to language $L$ iff ${\lambda} {\nu}  \in L \iff {\mu} {\nu} \in L$ for any string of input symbols ${\nu}$, where ${\lambda} {\nu} $ is a concatenation of ${\nu}$ after ${\lambda}$. The Myhill-Nerode theorem states that 
the number of states of the minimal DFA for  $L$ is the number of  {equivalence classes} of this
equivalence relation 
\cite{Hopcroft1979,Moore2019,nerode1958}.

\begin{figure}
    \centering
    \includegraphics[width=8cm]{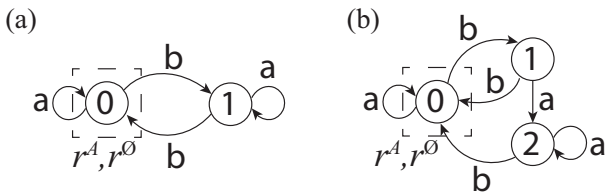}
    \caption{Two DFA that accept input strings with an even number of $b$s built from $\Lambda=\{a,b\}$. (a) The minimal DFA for this language; it is invertible. (b) A larger DFA that
    accepts the same language but is non-invertible; state 0 is a conflict state for $\rho_b$ and state 2 is a conflict state for $\rho_a$.}
    \label{fig:even_bs}
\end{figure}
Perhaps surprisingly, minimal LPDFA do not in general either maximise or minimise EP. This claim is our fifth main result;
to illustrate it, first consider
the two DFA in Fig.~\ref{fig:even_bs}, which
both have $\Lambda=\{a,b\}$ 
and accept input strings with an even number of $b$s. Fig.~\ref{fig:even_bs}\,(a) is the minimal DFA for this language. It is invertible, and 
so 
has zero EP.
% $\sigma_{\rho,q_{\mu}}(p(z^0_i,, {\lambda}_{-i}))=0$. 
% By contrast,
The larger DFA in
Fig.~\ref{fig:even_bs}(b) is non-invertible,
and so $\sigma^i(p(z^0_i, {\lambda}_{-i}))>0$ in general. For example, EP is positive if the sequences $(\Lambda_{i-2,} \Lambda_{i-1}, \Lambda_i)=(a\,{\rm or}\,b,b,a)$ and  $(\Lambda_{i-2,} \Lambda_{i-1}, \Lambda_i)=(b,a,a)$ both have non-zero probability. The minimal LPDFA never has higher EP than larger DFA, and often has lower EP.

\begin{figure}
    \centering
    \includegraphics[width=8cm]{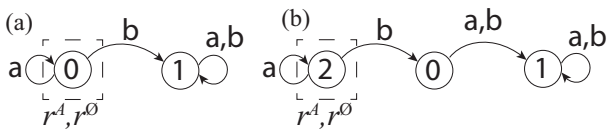}
    \caption{Two DFA that accept input strings built from an alphabet $\Lambda=\{a,b\}$ that contain no $b$s. (a) The minimal DFA that accepts this language. (b) An alternative, larger DFA.}
    \label{fig:no_bs}
\end{figure}

Now consider the two DFA in Fig.~\ref{fig:no_bs}. Both accept any input string constructed from $\Lambda=\{a,b\}$ with no $b$ symbols, and Fig.~\ref{fig:no_bs}(a) is the minimal DFA for this language. Neither DFA is invertible, so EP is generally non-zero for both. However, the non-minimal LPDFA in Fig.~\ref{fig:no_bs}\,(b) delays entropy production by a single iteration relative to Fig.~\ref{fig:no_bs}\,(a). As outlined in Section 5 of the Supplementary Information, this delay ensures that the overall EP for the larger LPDFA is always less than or equal to the EP for the minimal LPDFA.

\subsection{Languages are divided into costly and low-cost classes by the structure of their minimal DFA}

The DFA in Fig.~\ref{fig:no_bs}\,(b) can be extended, delaying non-zero EP. However, a finite number of additional states cannot prevent EP for arbitrary length inputs, and DFA are {\em necessarily} finite. Indeed, the sixth main result of our work, proven in detail in Section 6 of the Supplementary Information, is that if a minimal DFA is non-invertible, any DFA that accepts the same language must also be non-invertible. One cannot eliminate conflict states without disrupting the sorting of strings into equivalence classes. Thus 
if the minimal DFA for a regular language $L$ is non-invertible, recognising that language is inherently costly. Conversely, if the minimal DFA that accepts $L$ is invertible, recognising that language is low-cost.

As an example, consider a DFA that takes inputs of integers in base $n$, and accepts the integer $y$ if $y$ is divisible by $m$. As we show in Section 7 of the Supplementary Information, the minimal DFA for such a computation is invertible iff $n$ and $m$ have no common factors. Therefore, it is inherently costly to decide whether a number is divisible by 9 if the number is expressed in base 3, but not if the number is expressed in base 2, showing that even conceptually similar computations can have very different thermodynamic consequences.  

\section{Discussion}
\label{sec:discussion}

Breaking down complex computations into simple periodic updates, involving small parts of the computational system, is at the heart of both theoretical computer science and real-world computing devices. It is natural that physical systems designed to implement computations involve physical processes that are also local and periodic; that is how ``synchronous, clocked''
digital computers are designed.

However, physical systems that implement periodic, local computations are particularly vulnerable to stronger lower bounds on EP than the zero bound of the second law. Any physical operation -- including computations -- can, in principle, be performed in a thermodynamically reversible way,  with a sufficiently well-designed protocol \cite{Owen2019}. The nature of non-trivial computations, however, means that such a protocol would need to reflect not just the distribution of possible inputs to the computer, but also how those inputs are processed, and the subtle statistical coupling that is generated as the computation proceeds. 

We have illustrated how these challenges manifest as marginal and modularity mismatch costs in DFA with non-invertible local update maps. Interestingly, the overall computation performed by a DFA -- mapping the input word and starting computational state to the same input word and a final computational state -- is always invertible. The logical properties of the overall computation are therefore not helpful in understanding the necessary EP of a local, periodic device.

We have only a qualitative, system-specific understanding of
why the curves in \cref{fig:examp_ent_prodn}
and \cref{fig:priors} have the forms they do. Additionally, although similar results will hold for quantum mechanical or finite heat-bath treatments of DFA's thermodynamics, additional subtleties will arise.
More generally, DFA are just the simplest machine in the Chomsky hierarchy and it is unknown how marginal and modularity mismatch costs behave for other paradigms.
The constraints of locality and periodicity will also apply to 
(physical systems implementing) other machines in the hierarchy, 
such as push-down automata, RAM machines, or Turing Machines.
We would expect that 
variants of the
results concerning $\sigma_{\rm mod}$ and $\sigma_{\rm mar}$ presented here  also apply to those systems.
However, there will also be important differences. For example, the overwriting of input and/or memory that occurs in machines
more powerful than DFA will affect
$\sigma_{\rm mod}$ in ways not considered in this paper. Moreover, Turing machines and push-down automata have access to an infinite memory. DFA, by definition, do not -- indeed, it is this restriction that divides regular languages into low- and high-cost.

Finally, it is interesting to consider how the consequences of locality and periodicity relate to other resource costs. Recent work on transducers -- a computational machine that generates an output corresponding to a hidden Markov model -- has shown that a quantum advantage exists over a classical implementation if and only if the machine is not locally invertible \cite{Elliot2021}; it is unclear whether a similar result holds for DFA. The role of the input distribution in determining the thermodynamic costs in our work is also reminiscent of the way computational complexity depends on the distribution over inputs.

\hspace{2mm}

\noindent \textbf{Acknowledgements ---} TEO was supported by a Royal Society University Research Fellowship, and DHW was supported
by US NSF Grant CHE-1648973. DHW also thanks the Santa Fe Institute for support.

\hspace{2mm}

\noindent \textbf{Author contributions ---} TEO and DHW conceived the project. TEO performed the analysis. TEO and DHW interpreted the results and wrote the manuscript.

\hspace{2mm}

\noindent \textbf{Code availability ---} Code for calculating the data plotted in Fig.~\ref{fig:examp_ent_prodn}-\ref{fig:priors} can be found at https://doi.org/10.5281/zenodo.7950129.

\section{Methods}

\subsection{Relevant stochastic thermodynamics}
\label{sec:stoch_thermo}

\subsubsection{Mismatch cost}

Consider a system $X$ with a
finite set of states $\mathcal{X}=\{x_1,x_2,...\}$.
There is a distribution $p(x)$ over $\mathcal{X}$ at
some initial time, and that distribution evolves
according to a (potentially time-dependent) Markov process $\mu(t)$.
We assume that the system is attached to a single heat bath during this process,
choosing units so that the bath's temperature equals $1/k_B$. We also assume
that $\mu(t)$ obeys local detailed balance with respect to that bath and the system's
(potentially time-evolving) Hamiltonian \cite{Seifert2012}. Although we won't need to specify
whether the Markov process is discrete-time or continuous-time, to fix
the reader's intuition (and accord with real-world digital computers) we
can assume that it is continuous-time.

Suppose that the process runs for some pre-fixed time.
The distribution over $\mathcal{X}$
at the end of that time is a linear function of the initial distribution, which we write as
$
p^\prime(x^\prime) = \sum_x P(x^\prime | x) p(x)$, or just $p^\prime = Pp$ for short,
where $P$ is implicitly fixed by the stochastic process
$\mu(t)$.
A given $P$ will partition $\mathcal{X}$ into \textit{islands}. Two states $x$ and $x^\prime$ are within the same island if and only if  $P(x^{\prime \prime} | x) \neq 0$ and $P(x^{\prime \prime}|x^\prime) \neq 0$ for any state $x^{\prime \prime} $.

Let $q_\mu^c(x)$ be the initial probability distribution that minimizes the entropy production under $\mu(t)$  for distributions with support restricted to the island $c$.
This optimal distribution will be unique within each island.
No matter what the actual initial distribution $p$ is,
and regardless of the specific details of the process $\mu(t)$ that implements $P$,
so long as each $q^c_\mu$ has full support within island $c$,
the EP when the process is run with the initial distribution $p$ will be
\cite{Wolpert2020,Kolchinsky2021,Riechers2021}
\begin{equation}
    \sigma_{\mu}(p) = D(p \,||\, q_\mu) - D(Pp\,||\,Pq_\mu) + \sum_c p(c) \sigma_{\mu}(q_{\mu}^c).
    \label{eq:entropy_basic}
\end{equation}
Here, the index $c$ runs over the islands of the process, $p(c) = \sum_{x \in c} p(x)$  and $q_\mu(x) = \sum_c q(c) q^c_\mu(x)$ is called the  \textit{prior}  distribution~\cite{Wolpert2019stoch,Kolchinsky2020}. 

Note that the distribution over islands, $q_\mu(c)$, is arbitrary. Any distribution $q_\mu(x)$ that is a sum over the set of optimal distributions $\{q_\mu^c(x)\}$ 
could be used with the same results. In practice, the existence of many possible $q_\mu$ does not affect our analysis; we shall simply use a convenient $q_\mu$ with $q_\mu(c) \neq 0$ for all $c$.

The first two terms in Eq.~\ref{eq:entropy_basic} are the \textit{mismatch cost} 
\cite{Wolpert2020,Kolchinsky2021,Wolpert2019stoch} of the process. The final term in Eq.~\ref{eq:entropy_basic} is the \textit{residual entropy production}. Unlike the statistical mismatch cost, the residual EP depends on the physical details of the process
implementing $\mu(t)$. Each term in the sum is non-negative, but can be reduced to zero using a
quasi-static process \cite{Wolpert2020,Kolchinsky2021,Wolpert2019stoch}.

% \subsubsection{Modularity and marginal mismatch costs}
\subsubsection{Marginal and modularity mismatch costs}

Let $X_a$ and $X_b$ be two co-evolving systems
that are physically separated from one another 
during a time period $[0, 1]$, though they may have been coupled in the past. Due to this separation, we may consider separate protocols $\mu_a(t)$ and $\mu_b(t)$.  Moreover,
the prior for the overall process must be a product distribution, 
$q_\mu(x) = q_{\mu_a}(x_a) q_{\mu_b}(x_b)$
Taking $p(x_a)$ and $p(x_b)$ as the marginal distributions of the initial
joint distribution $p(x_a,x_b)$, the drop in KL divergence during
$[0, 1]$ is
\eq{
&D(p(x_a, x_b) || q_{\mu_a}(x_a) q_{\mu_b}(x_b)) \nonumber \\
& \qquad	- 
D( Pp(x_a, x_b) || P_a q_{\mu_a}(x_a) P_b q_{\mu_b}(x_b)),
}
where $P_a, P_b$
are the two matrices corresponding to the conditional distributions of ending states
given initial states.
This drop equals
\eq{
-\Delta H(p(x_a, x_b)) + \Delta H(p_a(x_a) || q_{\mu_a}(x_a)) \nonumber\\ + \Delta H (p_b(x_b) || q_{\mu_b}(x_b)),
}
where $H$ is the entropy, $H( . \,||\, .)$ is cross-entropy, and $\Delta$ means change from beginning to end of the evolution under $P$. Adding and subtracting marginal entropies, this form 
can be re-expressed as
\eq{
& -\Delta H(p(x_a, x_b)) + \Delta H(p_a(x_a)) + \Delta H(p_b(x_b)) \nonumber \\
& \qquad + \Delta D(p_a || q_{\mu_a}) + \Delta D (p_b || q_{\mu_b}).
}
By the definition of the change of mutual information between $X_a$ and $X_b$,  $\Delta I$, we obtain
\eq{
D(p(x_a, x_b) & || q_{\mu_a}(x_a) q_{\mu_b}(x_b)) = \nonumber \\
&\Delta D(p_a || q_{\mu_a}) + \Delta D (p_b || q_{\mu_b})  - \Delta I.
}
We may thus write
\eq{
\sigma_{\mu}(p) = \sigma_{\mu_a}(p_a) + \sigma_{\mu_b}(p_b) - \Delta I,
\label{eq:3a}
}
for the EP during $[0, 1]$, which simplifies to~\cref{eq:4a} if $X_b = X_{-a}$ and $X_{-a}$ is static. 

\cref{eq:4a} may, at first glance, seem inconsistent with the general discussion in Ref.~\cite{Wolpert2020}, which used a more general Bayes net formalism. In fact there is no inconsistency. In the language of Ref.~\cite{Wolpert2020},  the variables in $\mathbf{z}_0^i$ are the "parents" of $\mathbf{r}_i$, resulting in the same marginal and modularity mismatch costs as derived here.

% If $X_a$ is a subsystem of $X$, the dynamics of ${X}_a$ under $\mu(t)$ is said to be \textit{solitary} ~\cite{Wolpert2019stoch,Wolpert2020,Wolpert2020uncertainty} if $X_{-a}$ is static and the resulting heat flow is purely a function of the dynamics of ${X}_a$, with no dependence on $X_{-a}$. The mismatch cost of a solitary process
% can be split into two non-negative contributions \cite{Wolpert2019stoch}: a marginal mismatch cost of ${X}_a$ in isolation, in which all probabilities are marginalised over the degrees of freedom corresponding to ${X}_{-a}$, and a \textit{modularity} mismatch cost, given by the reduction in 
% mutual information between ${X}_a$ and $X_{-a}$ due to the dynamics induced by $\mu(t)$ \cite{Wolpert2019,Wolpert2020uncertainty, Boyd2018}.

% {\it Physical model.} 
\subsection{Physical model of DFA}
% \subsubsection{Assumptions in the model setup}

In order to apply stochastic thermodynamics to the computational model of DFA, it is necessary to make assumptions about how the logic is instantiated in a physical system. We assume that all the possible logical states of the system, defined by the set $R \times \Lambda^* \times \mathbb{Z}^+$ (combining the possible computational states, input words and iteration steps)
correspond to well-defined discrete physical states \cite{Seifert2012,Ouldridge2018Importance}. For example, the DFA could be a molecular assembly processing a copolymer tape~\cite{Brittain2021}. Metastable configurations of the assembly would represent the computational state, the sequence of the copolymer the state of the input word, and the position of the polymer the iteration. We also assume that if it is necessary to implement $\rho$, the DFA has access to ancillary hidden states -- which with probability $1$
are unoccupied at  the start and end of any update \cite{Owen2019}. 

Computation will, in general, involve an externally applied \textit{control protocol} that varies the physical conditions of
the system over time; in the case of the molecular computer, we would use time-varying concentrations of molecular fuel~\cite{Brittain2021}. This protocol defines the dynamics $\mu(t)$ discussed in Section~\ref{sec:stoch_thermo}. Although the dynamics will be stochastic, strictly speaking, we assume that $\mu(t)$ biases trajectories sufficiently to obtain effectively deterministic computation by the end of each update
More formally, we are interested in the limits of stochastic protocols under which they approximate deterministic dynamics to arbitrary accuracy\cite{Riechers2021impossibility}.
We abuse notation, using $\mu(t)$
to refer to both the external protocol and
the dynamics it induces over the system's states.

We take the input word $\boldsymbol{\lambda}$ to be a random variable  sampled from a distribution $p(\lambda)$. We use $\mathbf{r}_i$, $\mathbf{z}^0_i$, $\mathbf{z}^f_i$ and $\mathbf{c_i}$ to represent the random variables corresponding to the computational state of the DFA after update $i$, local state before and after update $i$; and the island occupied during iteration $i$, respectively.

We will consider a distribution $p(\lambda)$ in which all words are the same finite length $N$. Within this setup, a distribution of input words with lengths less than or equal to $N$ could be simulated by adding an extra null symbol that induces no computational transitions to the alphabet. Processing these null input symbols would have no thermodynamic cost under the assumptions considered here. For simplicity, we do not include these null symbols in our examples.

\subsection{Thermodynamic costs of DFA}
\subsubsection{Different measures of cost}
In this paper we focus on entropy production as the fundamental thermodynamic cost of running DFA. EP represents the lost ability to extract work from a system, and is a metric for the thermodynamic irreversibility. In certain contexts, the work required to perform a process, or the heat transferred to the environment therein, are also used to quantify the thermodynamic cost of a process.

The operation of a DFA does not increase the entropy of the computational degrees of freedom of the system, since the map from $(r_0 = r^\varnothing,\lambda)$ to $(r_N,\lambda)$ is one-to-one if the full input word is taken into account. If the computational states all have the same energy and intrinsic entropy \cite{Seifert2012,Ouldridge2018Importance} as is typically assumed, the energy and entropy change of the system will thus be zero. Any EP is equal to the heat transferred to the environment, which must be exactly compensated by the work done on the system. All three measures of thermodynamic cost are therefore identical.

\subsubsection{Costs considered in analysing the model}
We do not consider further the residual EP, nor the costs of incrementing $i$
(both can, in principle, be made arbitrarily small). We also neglect costs associated with actually generating $\mu(t)$ itself, as discussed in Ref.~\cite{Brittain2021}. Given these assumptions, whenever we use the term ``(minimal) EP'', we refer to the (minimal) EP due to the  mismatch cost (and its decomposition into marginal and modularity mismatch cost).

\subsubsection{Decomposition of EP generated at each iteration}

In general, when applying the mismatch cost formula to
a computation there are multiple choices for the times of the beginning
and end of the underlying process. This choice matters, because the mismatch cost contribution to EP is not additive over time. for example, the drop in KL divergence for a two-timestep computation will generally differ from the sum of the drops in KL divergence for each of those timesteps.

One could consider a single mismatch cost evaluated over the entire computation.
Under this choice, none of the details
of \textit{how} the conditional distribution $P$ of 
the overall computation 
arises by iterating the conditional distributions of each
step are resolved by the mismatch cost. All that matters is the drop in KL divergence between the
initial distribution, when the computer is initialized, and the ending distribution, when the output of the computation is determined.
This approach has been used to analyze Turing Machines \cite{Kolchinsky2020,Wolpert2019} as well as DFA~\cite{Kardes2022}.

An alternative choice is to focus on the EP generated at each iteration of the DFA, with the total EP of the entire computation being a sum of those iteration-specific EPs. 
Doing so allows us to manifest restrictions on the applied protocol inherent to the iterative process in the mismatch cost, rather than burying them in the residual entropy production of the computation as a whole. Given that we focus on costs arising from the iterative nature of the computation, it is natural to focus on the EP at each iteration of the DFA.

\bibliography{apssamp}% Produces the bibliography via BibTeX.

\end{document}

\preprint{APS/123-QED}

\title{Supplementary Information for ``Thermodynamics of deterministic finite automata operating locally and periodically''}% Force line breaks with \\

\author{Thomas E. Ouldridge}
 \email{t.ouldridge@imperial.ac.uk}
\affiliation{%
Imperial College Centre for Synthetic Biology and Department of Bioengineering, Imperial College London, London, SW7 2AZ, UK
}%

\author{David H. Wolpert}%
 \email{david.h.wolpert@gmail.com}
\affiliation{%
 Santa Fe Institute, Santa Fe, NM, 87501, USA \\
 Complexity Science Hub, Vienna, Austria \\
 Arizona State University, Tempe, AZ, USA \\
 International Center for Theoretical Physics, Trieste, Italy
}%

\date{\today}% It is always \today, today,

\maketitle

\newpage

\section{Simplification of $\sigma_{\rm mar}^i$ for LPDFA.}
\label{ap: simplify_sigma_local}

We first explicitly write the divergences as a sum over the islands and then a sum over states within islands: 
\begin{align}
\sigma^i_{\rm mar} & = D(p(z_i^0)\,\,||\,\,q_\mu(z_i^0)) - D(p(z_i^f)\,\,||\,\,q_\mu(z_i^f)) \nonumber \\
    &=
    \sum_{c_i}\sum_{z^0_i \in c_i} p(z^0_i|c_i) p(c_i)  \ln \frac{p(z^{0}_i|c_i)p(c_i)}{q_\mu(z^{0}_i|c_i) q_\mu(c_i)}
   \nonumber \\
    &- \sum_{c_i} \sum_{z^f_i \in c_i} p(z^f_i|c_i) p(c_i)  \ln \frac{p(z^f_i|c_i) p(c_i)}{q_\mu(z^f_i|c_i) q_\mu(c_i)} \nonumber\\
   & =
   \sum_{c_i} p(c_i) \sum_{z^0_i \in c_i} p(z^0_i|c_i)  \ln \frac{p(z^{0}_i|c_i)}{q_\mu(z^{0}_i|c_i)}
   \nonumber \\
   &+ \sum_{c_i} p(c_i) \ln \frac{p(c_i)}{q_\mu(c_i)} \nonumber \\
&    -
     \sum_{c_i} \sum_{z^f_i \in c_i} p(z^f_i|c_i) p(c_i)  \ln \frac{p(z^f_i|c_i) p(c_i)}{q_\mu(z^f_i|c_i) q_\mu(c_i)}.
    \label{eq:KLD1}
\end{align}
 Since the update deterministically collapses all probability within an island to a kronecker delta, $p(z^f_i|c_i) = q(z^f_i|c_i)$. Thus the final two terms in Eq.~\ref{eq:KLD1} cancel and we obtain
\begin{align}
 \sigma^i_{\rm mar} =
    & \sum_{c_i} p(c_i) D ( p(z^0_i|c_i)\,||\,q_\mu(z^0_i|c_i) ).
\end{align}

\section{Simplification of $\sigma^i_{\rm mod}$ for LPDFA.}
\label{ap:simplification_sigma_mod}

To calculate the modularity mismatch cost in an LPDFA,
it is helpful to separate 
$\boldsymbol{\lambda}_{j>i}$, the input string variables for $j > i$, from $\boldsymbol{\lambda}_{j<i}$, the variables for $j < i$. Making that separation 
% in the mutual information terms appearing 
% in Eq.~\ref{eq:EP1}, and
then using the chain rule for mutual information, we obtain
\begin{align}
    \sigma^i_{\rm mod} &= I(\mathbf{z}_i^0 ; \boldsymbol{\lambda}_{j<i}, \boldsymbol{\lambda}_{j>i}) - I(\mathbf{z}_i^f ; \boldsymbol{\lambda}_{j<i}, \boldsymbol{\lambda}_{j>i}) \nonumber \\
    & = I(\mathbf{z}_i^0 ; \boldsymbol{\lambda}_{j<i}) + I(\mathbf{z}_i^0 ; \boldsymbol{\lambda}_{j>i}| \boldsymbol{\lambda}_{j<i}) \nonumber \\
    & - I(\mathbf{z}_i^f ; \boldsymbol{\lambda}_{j<i}) - I(\mathbf{z}_i^f ; \boldsymbol{\lambda}_{j>i}| \boldsymbol{\lambda}_{j<i}),
\end{align}

Next, if we express the local state variable $\mathbf{z}_i$ in terms of the DFA's state variable and current input symbol's state variable, apply the chain rule again  and cancel terms, we get
\begin{align}
    \sigma^i_{\rm mod}
    & = I(\mathbf{r}_{i-1},\boldsymbol{\lambda}_i ; \boldsymbol{\lambda}_{j<i}) + I(\mathbf{r}_{i-1},\boldsymbol{\lambda}_i ; \boldsymbol{\lambda}_{j>i}| \boldsymbol{\lambda}_{j<i}) \nonumber \\
    & - I(\mathbf{r}_{i},\boldsymbol{\lambda}_i  ; \boldsymbol{\lambda}_{j<i}) - I(\mathbf{r}_{i},\boldsymbol{\lambda}_i  ; \boldsymbol{\lambda}_{j>i}| \boldsymbol{\lambda}_{j<i}) \nonumber \\
    & = I(\mathbf{r}_{i-1},\boldsymbol{\lambda}_i ; \boldsymbol{\lambda}_{j<i}) 
    + I(\mathbf{r}_{i-1} ; \boldsymbol{\lambda}_{j>i}| \boldsymbol{\lambda}_{j\leq i})
    \nonumber \\
    & - I(\mathbf{r}_{i},\boldsymbol{\lambda}_i  ; \boldsymbol{\lambda}_{j<i})  
    - I(\mathbf{r}_{i} ; \boldsymbol{\lambda}_{j>i}| \boldsymbol{\lambda}_{j\leq i}).
    \label{eq:condI}
\end{align}
Due to the deterministic and sequential operation of a DFA, both $\mathbf{r}_i$ and $\mathbf{r}_{i-1}$ are unambiguously determined by the first $i$ variables in the input string, $\boldsymbol{\lambda}_{j\leq i}$. As a result, 
the two conditional  information terms in the final line of Eq.~\ref{eq:condI} are both zero.
% \dhwc{This used to say ``both conditional  information terms in the final line of Eq.~\ref{eq:condI} are zero", but I don't think that's right, if the input symbols are not IID.} 
Applying the chain rule for mutual information twice to the remaining terms and simplifying, we obtain 
\begin{align}
    \sigma^i_{\rm mod}
    & =  I(\mathbf{r}_{i-1} ; \boldsymbol{\lambda}_{j<i}| \boldsymbol{\lambda}_i) 
    - I(\mathbf{r}_{i} ; \boldsymbol{\lambda}_{j<i}| \boldsymbol{\lambda}_i) \nonumber \\
        & =  I(\mathbf{r}_{i-1} ; \boldsymbol{\lambda}_{j\leq i}) - I(\mathbf{r}_{i-1} ; \boldsymbol{\lambda}_{i}) \nonumber \\
     &- I(\mathbf{r}_{i} ; \boldsymbol{\lambda}_{j\leq i}) + I(\mathbf{r}_{i} ; \boldsymbol{\lambda}_{i}).
    \label{eq:condI2}
\end{align}

Again using the fact that
the first $i$ variables in the input string, $\boldsymbol{\lambda}_{j \leq i}$, unambiguously specify both $\mathbf{r}_i$ and $\mathbf{r}_{i-1}$, $I(\mathbf{r}_{i-1} ; \boldsymbol{\lambda}_{j\leq i}) = H(\mathbf{r}_{i-1})$ and $I(\mathbf{r}_{i} ; \boldsymbol{\lambda}_{j \leq i}) = H(\mathbf{r}_{i})$. Thus, using the definition of the conditional entropy and mutual information,
\begin{align}
    \sigma^i_{\rm mod}
    & =  H(\mathbf{r}_{i-1}|\boldsymbol{\lambda}_i) - H(\mathbf{r}_{i}|\boldsymbol{\lambda}_i) \nonumber \\
    & = H(\mathbf{z}^0_i) - H(\mathbf{z}^f_i) \nonumber \\
      & = H(\mathbf{z}^0_i|\mathbf{c}_i) - H(\mathbf{z}^f_i|\mathbf{c}_i)
\label{eq:mod_mismatch_cost_penultimate}
\end{align}
where the last line follows by adding and subtracting $H(\mathbf{c}_i)$.

Finally, since the deterministic collapse of all inputs to a single output within an island ensures $H(\mathbf{z}^f_i|\mathbf{c}_i)=0$, we can further
reduce the modularity mismatch cost to
\begin{align}
    \sigma^i_{\rm mod}
    & = 
H(\mathbf{z}^0_i|\mathbf{c}_i)
    \label{eq:condI3app}
\end{align}
This result establishes the claim made in the main text.

\section{Estimating entropy production for an uncertain input distribution.}
\label{ap:uncertain}

The designer's best estimate for the entropy production is obtained by averaging Eq.~13 of the main text over $\pi(v)$ and $\pi(w|v)$:
% \begin{equation}
\begin{eqnarray}
\hat{\sigma}^i 
= \sum_{v, c_i, w} \pi(v)  p_v(c_i) \pi(w|v)  
H\left( p_w(z_i^0 |c_i) \,||\, q_\mu(z_i^0 |c_i) \right).
\label{eq:KLD5}
\end{eqnarray}
% \end{equation}
Expanding the cross entropy yields 
\begin{align}
&\hat{\sigma}^i \nonumber\\
=& -\sum_v \pi(v) \sum_{c_i} p_v(c_i) \sum_w \pi(w|v) \sum_{z_i^0 \in c_i}p_w(z_i^0 |c_i) \ln q_\mu(z_i^0 |c_i) \nonumber
\\
=&\sum_v \pi(v) \sum_{c_i} p_v(c_i) \nonumber \\ 
&\sum_{z_i^0 \in c_i} \sum_w \pi(w|v) p_w(z_i^0 |c_i) \ln \frac{\sum_w \pi(w|v) p_w(z_i^0 |c_i)}{q_\mu(z_i^0 |c_i)} \nonumber
\\
-& \sum_v \pi(v) \sum_{c_i} p_v(c_i) \nonumber \\
&\sum_{z_i^0 \in c_i} \sum_w \pi(w|v) p_w(z_i^0 |c_i) \ln {\sum_w \pi(w|v) p_w(z_i^0 |c_i)} \nonumber\\
=&\sum_v \pi(v) \sum_{c_i} p_v(c_i) D( \hat{p}(z_i^0 |c_i,v)  \,||\,  q_\mu(z_i^0 |c_i)) \nonumber \\
-& \sum_v \pi(v) \sum_{w} \pi(w|v) \sum_{c_i} p_v(c_i) 
\sum_{z_i^0 \in c_i} p_w(z_i^0 |c_i) \ln p_w(z_i^0 |c_i)\nonumber
\\
& \; - \sum_v \pi(v) \sum_{c_i} p_v(c_i) 
\sum_{z_i^0 \in c_i} \hat{p}(z_i^0 |c_i) \ln {\hat{p}(z_i^0 |c_i)}  \nonumber
\\
+& \sum_v \pi(v) \sum_{w} \pi(w|v) \sum_{c_i} p_v(c_i) 
\sum_{z_i^0 \in c_i} p_w(z_i^0 |c_i) \ln p_w(z_i^0 |c_i)\nonumber
\\
= & H(\mathbf{z}_i^0|\mathbf{c}_i,\mathbf{v},\mathbf{w})\nonumber\\
+&\sum_v \pi(v) \sum_{c_i} p_v(c_i) D( \hat{p}(z_i^0 |c_i,v)  \,||\,  q_\mu(z_i^0 |c_i)) \nonumber \\
+& I( \mathbf{z}_i^0;\mathbf{w}|\mathbf{c}_i,\mathbf{v}) .  
%&- \sum_y \pi(y) \sum_{c_i} p_y(c_i) \sum_{z_i^0 \in c_i} \sum_x \pi(x|y) p_x(z_i^0 |c_i) \ln {\sum_x \pi(x|y) p_x(z_i^0 |c_i)} \nonumber\\
%&\sum_x \pi(x|y)   \nonumber\\
%& =  \sum_{z_i^0}  \sum_x  \pi(x|y) p_x(z_i^0 |c_i) \ln \frac{p_x(z_i^0 |c_i)}{q_\mu(z_i^0 |c_i)}\nonumber\\
%& = \sum_{z_i^0}  \sum_x \pi(x|y)  p_x(z_i^0 |c_i)  \ln \frac{\sum_x \pi(x|y) p_x(z_i^0 |c_i)}{q_\mu(z_i^0 |c_i)} \nonumber\\
%&- \sum_{z_i^0}  \sum_x  \pi(x|y)  p_x(z_i^0 |c_i)  \ln \sum_x \pi(x|y) p_x(z_i^0 |c_i) \nonumber\\
%& + \sum_{z_i^0}  \sum_x  \pi(x|y)  p_x(z_i^0 |c_i)  \ln \pi(x|y) p_x(z_i^0 |c_i) \nonumber\\
%& - \sum_{z_i^0}  \sum_x  \pi(x|y)  p_x(z_i^0 |c_i)  \ln \pi(x|y) \nonumber\\
%& = D\left( \left(\sum_x \pi(x|y){p}_x(z_i^0 |c_i)\right)  \,||\,  q_\mu(z_i^0 |c_i)\right) \nonumber \\
%&+ \left(H \left[\sum_x \pi(x|y){p}_x(z_i^0 |c_i)\right] + H[\pi(x|y)] -H[\pi(x|y)p_x(z_i^0|c_i)] \right). \nonumber \\
\label{eq:KLD6app}
\end{align}
Here, $H(\mathbf{z}_i^0|\mathbf{c}_i,\mathbf{v},\mathbf{w})$ and $I( \mathbf{z}_i^0;\mathbf{w}|\mathbf{c}_i,\mathbf{v})$ are defined with respect to the joint distribution estimated by the designer, $\hat{p}(v,w,z_i^0,c_i)= \pi(v)\pi(w|v)p_v(c_i)p_w(z_i^0 |c_i)$, and $\hat{p}(z_i^0 |c_i,v) = \sum_w \pi(w|v){p}_w(z_i^0 |c_i) $ is  the designer's estimate for the probability distribution within an island, having averaged over  $\pi(w|v)$.

\section{Minimal DFA are not necessarily more thermodynamically efficient than larger DFA.}
\label{ap:entropy_delay}
We claim that, for any distribution of inputs and choice of iterated protocol for the LPDFA in Fig.~7\,(a) of the main text, it is possible to choose a protocol for the LPDFA in Fig.~7.\,(b) that results in EP that is less than or equal to the EP of the LPDFA in  Fig.~7\,(a). To prove this claim, note that the EP at iteration $i$ for the minimal DFA in  Fig.~7\,(a) is given by considering only the island defined by $\{(b,0),(b,1)\}$. Thus 
\begin{align}
   \sigma^i(p(z_i^0, {\lambda}_{-i})) &= p_i(b,0) \ln q_{\mu}(b,0|b,0\,{\rm or}\,1) \nonumber \\&+p_i(b,1) \ln q_{\mu}(b,1|b,0\,{\rm or}\,1).
\end{align}

For the larger LPDFA in in  Fig.~7\,(b), the EP at iteration $i$ is entirely due to the two islands defined by $\{(b,0),(b,1)\}$ and $\{(a,0),(a,1)\}$. Thus
\begin{align}
   {\sigma^i}^\prime(p^\prime(z^0_i,{\lambda}_{-i})) &= p^\prime_i(b,0) \ln q^\prime_{\mu}(b,0|b,0\,{\rm or}\,1) \nonumber \\
   &+p^\prime_i(b,1) \ln q^\prime_{\mu}(b,1|b,0\,{\rm or}\,1) \nonumber \\
   &+p^\prime_i(a,0) \ln q^\prime_{\mu}(a,0|a,0\,{\rm or}\,1) \nonumber \\
   &+p^\prime_i(a,1) \ln q^\prime_{\mu}(a,1|a,0\,{\rm or}\,1),
\end{align}
with primed quantities referring to the larger DFA for clarity. 
Given the  well-defined starting state of the LPDFA, it is possible to say that none of  these states are occupied at the first step: $p^\prime_1(b,0)=p^\prime_1(a,0)=p^\prime_1(b,1)=p^\prime_1(a,1)=0$. Moreover, assuming the same distribution of input strings to both devices, the related structure of both devices implies $p_i(b,0) = p^\prime_{i+1}(b,0)+p^\prime_{i+1}(a,0)$ and $p_i(b,1) = p^\prime_{i+1}(b,1)+p^\prime_{i+1}(a,1)$. If we then chose protocols for the larger DFA so that $q^\prime_{\mu}(b,0|b,0\,{\rm or}\,1) = q^\prime_{\mu}(a,0|a,0\,{\rm or}\,1)=q_{\mu}(b,0|b,0\,{\rm or}\,1)$ and $q^\prime_{\mu}(b,1|b,0\,{\rm or}\,1) = q^\prime_{\mu}(a,1|a,0\,{\rm or}\,1)=q_{\mu}(b,1|b,0\,{\rm or}\,1)$, we obtain
\begin{align}
   &{\sigma^i}^\prime(p^\prime(z^0_i,{\lambda}_{-i}))) \\
   =& \left(p^\prime_i(b,0)+p^\prime_i(a,1) \right) \ln q_{\mu}(b,0|b,0\,{\rm or}\,1) \nonumber \\
   +& \left(p^\prime_i(a,1)+p^\prime_i(b,1)\right) \ln q_{\mu}(b,1|b,0\,{\rm or}\,1) \nonumber \\
   =&\sigma^{i-1}(p(z^0_{i-1},{\lambda}_{-(i-1)}))
\end{align}
for $i>1$, and ${\sigma^i}^\prime(p^\prime(z^0_i,),{\lambda}_{-i})=0$ for $i=1$. As a result, for any finite number of iterations $N$, 
\begin{align}
&\sum_{i=1}^N \sigma^i(p(z^0_i),{\lambda}_{-i}) -\sum_{i=1}^N {\sigma^i}^\prime(p^\prime(z^0_i,{\lambda}_{-i})) \\ &=\sigma^i(p(z^0_N),{\lambda}_{-N}) \geq 0.
\end{align}

\section{If a minimal DFA is non-invertible, so is any DFA that accepts the same language}
\label{ap:minimal_non_invertible}

To prove the claim, recall that a non-invertible DFA has at least one ``conflict state" to which multiple input computational states are mapped by the same input symbol under $\rho$ (see Fig.~5 of the main text). We consider the network $\rho_\lambda$, defined by the mapping between computational states for an input symbol $\lambda$ corresponding to such a conflict state in a minimal DFA $D_L$ that accepts the language $L$. If $D_L$ has $M$ computational states, there are exactly $M$ directed edges in $\rho_\lambda$. Thus the existence of a conflict state with more than one inward edge implies at least one state with zero inward edges. The existence of such a state $r^\dagger$ in $\rho_\lambda$ implies that there are no transitions into the equivalence class represented by $r^\dagger$ due to the symbol $\lambda$. 

The states in any other DFA $D^\prime_L$ that accepts $L$ can be partitioned into non-overlapping sets, each of which corresponds to an equivalence class of $L$ (one of the states of $D_L$ -- see Refs. [26-27] of the main text). The transitions between these non-overlapping sets must exactly match the transitions defined by $\rho$ in $D_L$, otherwise $D^\prime_L$ would fail to sort input strings into equivalence classes of $L$. Therefore, if $r^\dagger$ has no inward edges in the network $\rho_\lambda$ defined by $D_L$, none of the states in the set corresponding to the equivalence class represented by $r^\dagger$ can have inward edges in the network $\rho^\prime_\lambda$ defined by $D^\prime_L$. The existence of at least one state in the network $\rho^\prime_\lambda$ with zero inward edges implies the existence of at least one conflict state with two or more inward edges in $\rho^\prime_\lambda$, since the total number of edges is equal to the total number of states. Therefore any $D^\prime_L$ that accepts the same language as a non-invertible minimal DFA $D_L$ must exhibit conflict states, and must also be non-invertible. 

\section{The invertibility of DFAs that accept words in base $n$ that are divisible by $m$}
\label{ap:factors}
In the context of these DFA, it is helpful to refer to the alphabet using numerical indeces. We assume that the integer $y$ is written on the tape in base $n$ so that its most significant figure is ${{\lambda}_1}$, its second most significant figure is ${{\lambda}_2}$, {\it etc}.

Let $y_i$ be the integer represented by the first $i$ entries in the input word. The DFA will be in the absorbing state $r^A$ after iteration $i$ if and only if $y_i\,\textrm{mod}\,{m} = 0$. Moreover, after the next iteration, the system will be in $r^A$ if and only if 
\begin{equation}
y_{i+1}\,\textrm{mod}\,{m} = \left(\lambda_{i+1} + n(y_i\,\textrm{mod}\,{m})\right) \,\textrm{mod}\,{m} = 0.
\label{eq:mod_equality1}
\end{equation}
The value of $y_i \,\textrm{mod}\,m$ is thus sufficient to express the specify the equivalence class of the word fragment $y_i$, since it is the only information needed from the first $i$ digits to determine whether the full word is divisible by $m$. We note, however, that knowledge of $y_i$ is not {\em necessary} to specify the equivalence class. In general, 
words with distinct values of $y_i \,\textrm{mod}\,m$ can belong to the same equivalence class. Nonetheless, the equivalence class corresponding to the absorbing state necessarily only contains word fragments with $y_i \,\textrm{mod}\,m=0$.

\subsection{$m$ and $n$ have no common factors}
 Due to the arguments in Section~\ref{ap:minimal_non_invertible} of the SUpplementary Information, it is sufficient to show that any DFA that accepts this language is invertible. We may therefore consider a DFA in which there are $m$ states, each one corresponding to a single value of $y_i \,\textrm{mod}\,m$. Let us assume, for the sake of contradiction, that such a DFA is non-invertible. For this to be true, we require that two distinct values of $y_i \,\textrm{mod}\,m$, which would lead to different computational states after iteration $i$, result in the same value of $y_{i+1} \,\textrm{mod}\,m$ for a given $\lambda_{i+1}$. Using the expression for $y_{i+1} \,\textrm{mod}\,m$ in Eq.~\ref{eq:mod_equality1}
\begin{equation}
\left(\lambda_{i+1} + n k \right) \,\textrm{mod}\,{m} = \left(\lambda_{i+1} + n l \right) \,\textrm{mod}\,{m},
\label{eq:mod_equality2}
\end{equation}
where $l, k$ are two distinct integers between 0 and $m-1$. We will assume $k>l$ without loss of generality.

Using the properties of modular arithmetic, we may rewrite Eq.~\ref{eq:mod_equality2} as
\begin{equation}
\left( n (k-l) \right) \,\textrm{mod}\,{m} = 0.
\label{eq:mod_equality3}
\end{equation}
Eq.~\ref{eq:mod_equality3} can only be zero if $k-l$ is zero, which would violate the requirement that $l\neq k$, or if the union of the prime factors of $n$ and $k-l$ is a superset of the prime factors of $m$. Since $k-l<m$, the prime factors of $l-k$ alone cannot be a superset of the prime factors of $m$. Therefore $n$ and $m$ must share at least one prime factor, violating the initial assumption and proving the claim by contradiction.

\subsection{$m$ and $n$ have at least one common factor}
We now prove that the minimal DFA that accepts words written in base $n$ that are divisible by $m$ is non-invertible if $n$ and $m$ have at least one common factor. To do so, it is sufficient to show that at least one non-zero value of $y_i \,\textrm{mod}\,m$ results in $y_{i+1} \,\textrm{mod}\,m = 0$ for $\lambda_{i+1}=0$, since this operation corresponds to a non-accepting state being mapped to $r^A$ by $\lambda_{i+1}=0$, and $r^A$ will also be mapped to $r^A$ by $\lambda_{i+1}=0$. In other words, we require
\begin{equation}
\left(0 + n k \right) \,\textrm{mod}\,{m} = 0
\label{eq:mod_equality4}
\end{equation}
for integer $k$, $0<k<m$. For any $n,m$ that share a common factor $g$, there will always be a $k=m/g$ for which Eq.~\ref{eq:mod_equality4} holds. Thus any DFA that accepts words written in base $n$ that are divisible by $m$ will be non-invertible if $n$ and $m$ have at least one common factor.